\documentclass[11pt,a4paper]{article}
\pdfoutput=1
\usepackage{graphicx}

\usepackage{jheppub}

\usepackage{color}

\usepackage{ulem}
\usepackage{psfrag}
\usepackage{hyperref}
\newcommand{\Eqref}[1]{eq.~\eqref{#1}}

\title{Can we see quantum gravity?\\ Photons in the asymptotic-safety scenario}

\author[a,1]{Babette D\"obrich}
\affiliation[a]{Theoretisch-Physikalisches Institut, Max-Wien-Platz 1, Friedrich-Schiller-Universit{\"a}t
Jena,\\
\& Helmholtz Institut Jena, Fr\"obelstieg 3, D-07743 Jena, Germany
\note{Now at Deutsches Elektronen-Synchrotron DESY, Notkestra\ss e 85, 22607 Hamburg, Germany}}

\author[b]{Astrid Eichhorn}
\affiliation[b]{ Perimeter Institute for Theoretical Physics, 31 Caroline Street N, Waterloo, N2L 2Y5, Ontario, Canada}

 \emailAdd{babette.doebrich@desy.de}
 \emailAdd{aeichhorn@perimeterinstitute.ca}

\preprint{}             

\abstract{In the search for a quantum theory of gravity it is crucial to find experimental access to quantum gravitational effects. Since these are expected to be very small at 
observationally accessible scales it is advantageous to consider processes with no tree-level contribution in the standard model, such as photon-photon scattering. 
We examine the implications of asymptotically safe quantum gravity in a setting with extra dimensions for this case, and point out that various near-future photon-collider setups, employing either electron or muon colliders, or even a purely laser-based setup, could provide a first observational 
window into the quantum gravity regime.
}


\begin{document}

\maketitle

\section{Introduction}

The search for the quantum gravity theory which is realised in our universe is a very challenging task. Several competing candidate theories exist, which are currently only restricted by the 
requirement of internal consistency, and the necessity to recover Einstein gravity at large distances. This freedom has allowed many scenarios to develop, which are founded on 
very different assumptions about the fundamental nature of spacetime, the microscopic gravitational degrees of freedom, and the realisation of fundamental symmetries.
Thus it is mandatory to find experimental or observational tests, which will provide further guidance in the construction of the theory of quantum gravity. Here, we consider the process of 
photon-photon scattering  in asymptotically safe quantum gravity to discuss possible experimental tests. 

Photon-photon scattering is a very special process, arising purely from quantum effects, without any contribution from classical physics, tree-level 
contributions being absent within the Standard Model. 

At the eV-scale, which is the relevant scale for optical setups, its main contribution arises from an electron-loop diagram
\cite{Euler:1935zz,Karplus:1950zz}, which yields a small contribution $\sim \alpha^4$
with the QED-coupling 
$\alpha \approx \frac{1}{137}$ at these scales, see, e.g.
\cite{Dittrich:2000zu}\footnote{At $\omega \ll m_e$ the factor $\omega^6/m_e^8$, with the electron mass $m_e$ and photon energy $\omega$
yields an additional suppression.}. 
An experimental confirmation of 
this process is thus not yet available, but near-future laser-based experiments will soon reach the necessary energy scales, see, e.g. \cite{Heinzl:2006xc,Marklund:2008gj,King:2012aw}. 
\\
At energies well above the GeV scale, photon-photon interaction is dominated by weak-interaction processes, and receives its main contribution from a W-boson loop. 
This yields the background to the quantum gravity contribution which we will discuss here.\\
Since no tree-level-contribution to photon-photon scattering exists in the Standard Model, it is particularly sensitive to new 
physics at the optical scale \cite{Gies:2008wv} as well as in the x- and gamma-ray regimes 
\cite{Ringwald:2001cp}. The latter energy scales might also provide 
insight into quantum gravity effects.

In this work we assume that the metric carries the quantum gravitational degrees of freedom, and gravity can be quantised in a standard path-integral framework 
based on a continuous nature of spacetime. This assumption certainly holds within an effective-field-theory treatment 
of quantum gravity, at scales presumably below the Planck scale, see, e.g. \cite{Donoghue:1993eb,Burgess:2003jk}. 
Contrarily to what one might think from a perturbative point of view, such an assumption even allows to construct a UV completion for gravity:
Within asymptotically safe quantum gravity \cite{Weinberg:1980gg}, the metric carries the dynamical degrees of freedom of gravity up to arbitrarily 
high momentum scales. As is well known, perturbative renormalisability fails for this theory \cite{'tHooft:1974bx,Goroff:1985th}. 
A finite and predictive theory can nevertheless be constructed in a non-perturbative setting. It requires the existence of a non-Gau\ss{}ian fixed point in the $\beta$ functions of the 
dimensionless running couplings: This scenario works in an analogous way to renormalisability in non-Abelian Yang-Mills theories: 
There, the theory becomes asymptotically free at large energies. In the language of the functional Renormalisation Group this statement 
translates into the fact that the theory has a UV-attractive Gau\ss{}ian, i.e. non-interacting fixed point. At the fixed point, the theory 
becomes scale-free, and thus it is possible to take the limit of infinite momentum cutoff while keeping physical quantities finite. \\
Similarly, 
a non-Gau\ss{}ian, i.e. interacting fixed point which is approached at high momenta, allows to take the infinite-cutoff limit while 
arriving at finite physical predictions \cite{Weinberg:1980gg}.

These ideas present a generalisation of the perturbative concepts of renormalisability, and can apply to any quantum field theory, not just quantum gravity, 
see, e.g. \cite{Gies:2009hq,Gies:2009sv,Braun:2010tt,Fabbrichesi:2010xy}.
\\
The non-trivial, and technically challenging part of this proposal is 
that it cannot straightforwardly be tested within perturbation theory. The fact that the UV theory is interacting implies that the fixed 
point is accessible with non-perturbative means. A fully non-perturbative formulation of the functional Renormalisation 
Group \cite{Wetterich:1993yh}, for reviews see \cite{Berges:2000ew,Polonyi:2001se,Pawlowski:2005xe,Gies:2006wv,Rosten:2010vm}, has 
allowed to collect a 
substantial amount of evidence for the existence of the fixed point \cite{Reuter:1996cp,Dou:1997fg,Lauscher:2001ya,Reuter:2001ag,
Lauscher:2002sq,Litim:2003vp,Lauscher:2005xz,Fischer:2006fz,Codello:2008vh,Benedetti:2009rx,
Eichhorn:2009ah,Groh:2010ta,Eichhorn:2010tb,Manrique:2010am,Manrique:2011jc,Donkin:2012ud}, and its compatibility with Standard 
Model matter \cite{Percacci:2002ie,Percacci:2003jz, Eichhorn:2011pc, Eichhorn:2011ec}, even allowing for a possibility to solve the triviality 
problem in QED and the Higgs sector \cite{Harst:2011zx,Zanusso:2009bs,Shaposhnikov:2009pv}, for reviews see \cite{AS_reviews}. 
For further indications for the existence of the fixed point, see also 
\cite{Smolin:1981rm,Christensen:1978sc,Gastmans:1977ad,Niedermaier:2010zz,Ambjorn:2010rx}.

Since this proposal for quantum gravity allows to construct a quantum field theory for the metric, all QFT-tools are at our disposal in order to analyse 
the quantum gravitational contribution to photon-photon scattering. We use a scale-dependent version of the effective action to evaluate the leading-order contributions to this 
process. The RG-scale dependent effective action contains the effect of all quantum fluctuations in the path-integral, which have momenta larger than the RG momentum scale $k$. 
Thus a tree-level evaluation of the effective action suffices to evaluate the leading-order contribution to physical processes such as photon-photon scattering. The non-trivial 
information on the scale-dependence is carried by the $k$-dependent coupling constants, in this case the Newton coupling $G_N(k)$.
In particular,
a tree-level scattering 
amplitude via an internal graviton exists, which constitutes the leading contribution to this process (for comments on the effect of further operators in the effective action 
see sect.~\ref{access_Mstar}). 

Let us emphasise the difference between asymptotic safety and an effective field theory approach with a UV cutoff: In the latter, photon-photon scattering via an internal graviton also occurs, but the cross section diverges in the limit of large momenta, and violates perturbative unitarity. 
Within asymptotic safety we find that the corresponding process yields a cross section that does not diverge as a function of the momentum, 
and satisfies the bound $\sigma \sim \frac{1}{p^2}$ at high momenta. 
 This is due to the 
fact that the scenario relies on the existence of an interacting fixed point for the dimensionless couplings in the far UV. Consequently, the dimensionful Newton 
coupling becomes a running coupling, and at high momenta $k$ scales as
\begin{equation}
G_N(k^2)= \frac{G_{\ast}}{k^2},\label{runningNewton}
\end{equation}
where $G_{\ast}= \rm const$ is the finite, dimensionless fixed-point value of the coupling.
This behaviour of the dimensionful Newton coupling implies that the cross section will be well-behaved at high energies, see \cite{Weinberg:1980gg}. 

Note the following subtlety here: The momentum scale $k$ in the scaling relation for the Newton 
coupling is an RG scale, meaning that it denotes the scale of an effective theory, where the high-momentum fluctuations in the path integral have been integrated out. 
The identification of this RG scale with a physical momentum scale is only straightforwardly possible in single-scale problems, where it is meaningful to use a tree-level evaluation of the 
scale-dependent effective action to access physical processes. Within our setting, where the center-of-mass energy $E_{\rm cm}$ is the only relevant momentum scale of the problem, the 
identification of the RG scale and the physical momentum scale appears reasonable. 

Let us point out the difference between our treatment and a perturbative approach, where metric fluctuations are linearised around a flat background in perturbation theory: 
In our case, we also obtain a graviton propagator from linearising the action around an appropriate background, which for our experimental setting is Minkowski spacetime. 
The crucial difference is that our results rely on the non-perturbative information that is encoded in the running coupling $G_N(k)$ in the effective theory: 
The behaviour of $G_N(k)$ is obtained from a fully non-perturbative evaluation of the path-integral, and contains the effect of arbitrarily large metric fluctuations. 
Thus our results carry the fully non-perturbative behaviour of asymptotically safe quantum gravity.

Let us now start by evaluating photon-photon scattering in this setting, first in a setting with $3+1$ dimensional Minkowski spacetime, and then in a generalised context 
of $n$ extra dimensions.

\section{Photon-photon scattering in asymptotically safe quantum gravity}

 \begin{figure}[!here]
\begin{center}
 \includegraphics[width=0.7\linewidth]{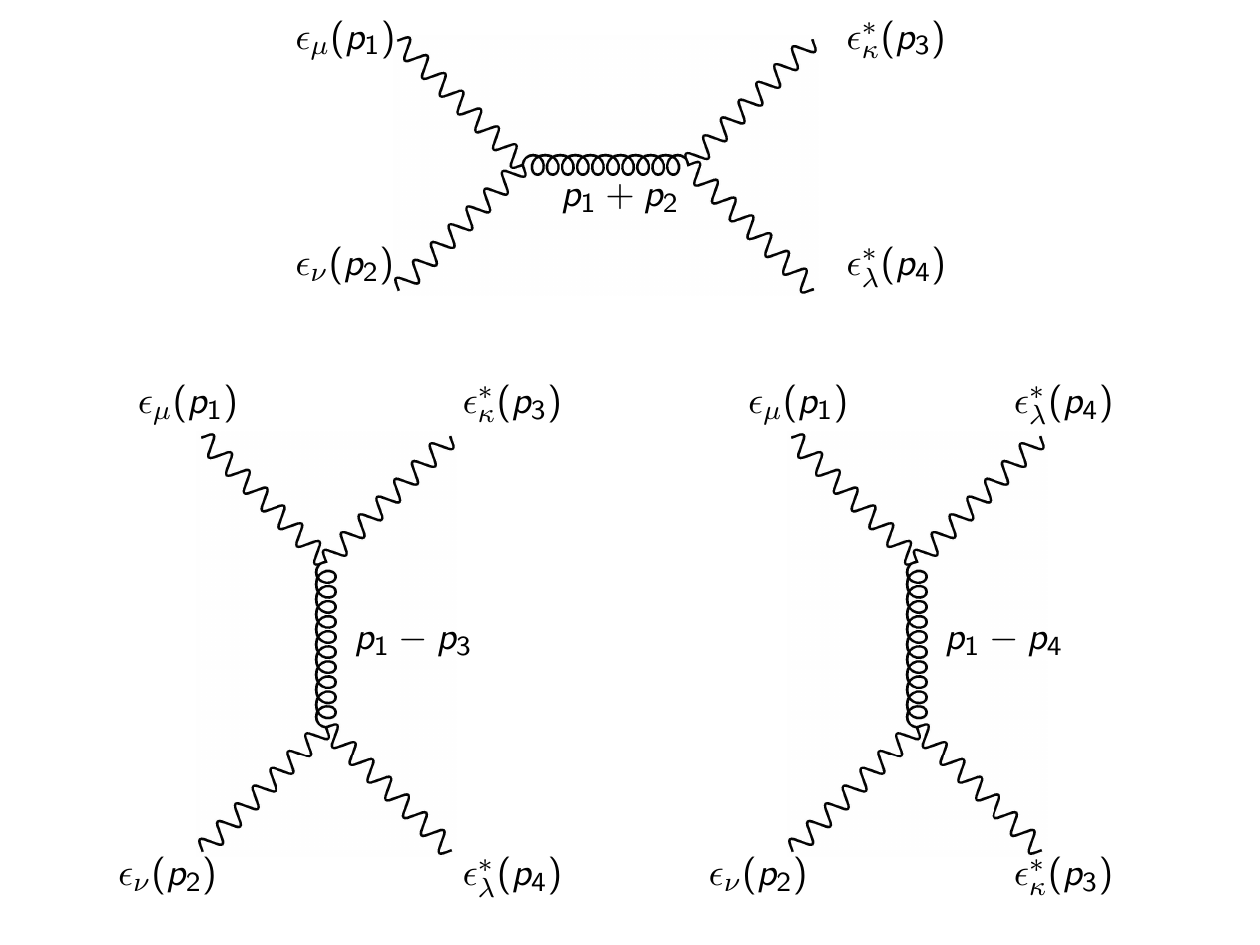}
\end{center}
\caption{ $s-$, $t-$ and $u-$channel diagrams for photon-photon scattering through graviton exchange. Photons are depicted by wiggly, gravitons by curly lines.
$\epsilon_\alpha$ denotes the polarisation of the photons. \label{diags}}
\end{figure}
At tree-level, an $s-$, $t-$ and $u-$channel diagram contribute to the amplitude for $2 \gamma \rightarrow 2 \gamma$, see fig.~\ref{diags}. To arrive at the unpolarised 
cross-section, we need to evaluate $\frac{1}{4}\sum_{\lambda_1,\lambda_2,\lambda_3,\lambda_4}\vert\mathcal{M}_{\lambda_1\lambda_2\lambda_3\lambda_4}\vert^2$, where $\mathcal{M}$ denotes 
the total amplitude for a specific choice of polarisations $\lambda_i$, and the factor $1/4$ is due to the polarisation average.

For clarity, we state the differential cross section for the process depicted in fig. \ref{diags}, in the case $d=4$, see  \cite{Gupta}
\begin{equation}
\frac{d \sigma}{d \Omega}=32 E^2 G_N^2(E) \csc(\theta )^4 \left[1+\cos\left(\frac{\theta }{2}\right)^{16}+\sin\left(\frac{\theta }{2}\right)^{16}\right], \label{Gupta}
\end{equation}
where we have identified the RG scale with the energy $E= \sqrt{s}/2$ in $G_N(E)$.
For details of the calculation and the general result in $d$ dimensions see appendix \ref{sec:app_calc}.
Note that the t- and u-channels contribute a singularity at $\theta=0$, as the virtual graviton goes on shell. This property is well known from scattering cross sections 
involving massless intermediate fields, as, e.g. Coulomb scattering.

The crucial new point in this work is to include the effect of a momentum-dependent Newton coupling, which encodes the non-perturbative physics underlying the asymptotic-safety scenario. 
Let us explain how we implement this scale-dependence:
As is well known, the running of Newton's coupling is non-universal in $d=4$, 
since it is a dimensionful coupling. Nevertheless, there are two general properties that the scale-dependence of Newton's coupling has to satisfy within the asymptotic-safety 
scenario: The first is, that $G_N = {\rm const}$ on a large range of scales from the sub-mm regime to astrophysical distances. This behaviour characterises 
the classical regime of the theory.
It holds irrespective of the UV completion for gravity, and for the case of asymptotic safety it has been shown that the $\beta$ functions admit 
solutions which satisfy this experimental requirement \cite{Reuter:2004nx}. 
The second property pertains only to the asymptotic-safety scenario: A UV completion with the help of an interacting fixed point requires that $G(k) \rightarrow G_{\ast}$ for 
$k \rightarrow \infty$, where $G(k)=G_N(k)k^{d-2}$ is the dimensionless Newton coupling. Accordingly we deduce that $G_N(k) \sim \frac{1}{k^2}$
 at high momenta. The value $G_{\ast}$ 
itself is non-universal, but the scaling property is universal. Thus we model the scale-dependence in a very simple way, which makes use only of these two properties:
\begin{equation}
G_N(k) = \Bigg\{ 
\begin{array}{c c}
G_{\rm Newton} &\mbox{ for } k^2 < k_{\rm trans}^2\\
G_{\rm Newton}\frac{k_{\rm trans}^2}{k^2}& \mbox{ for } k^2 \geq k_{\rm trans}^2
\end{array},\label{runningG}
\end{equation}
where $G_{\rm Newton}$ is the constant value for the Newton coupling that we measure in the infrared. Replacing $G_N(k)$ by a function that smoothly interpolates between the two limits is possible and does not affect our results strongly, since evaluations of truncated RG flows indicate a fast transition between the two regimes, see, e.g. \cite{Gerwick:2011jw}.
Here the transition scale $k_{\rm trans}$ is presently unknown, and signals the onset of the fixed-point behaviour. One might conjecture, that this scale is close to 
the Planck scale, however this is by no means clear. In fact, the transition scale might lie considerably below or above the Planck scale.
Thus one of the goals of a 
phenomenology of asymptotic safety must be the determination of this scale. Let us stress that this is not a question of increasing the precision of calculations, in fact no such scale can in principle be determined from an RG flow, but is to be determined from observations. \\

Let us comment on the important question of how to identify the RG scale $k^2$ with a physical scale of the process under consideration. The first crucial issue here 
is that whenever one considers a {\it free} particle, it is not correct to identify its three-momentum or energy with the RG-scale $k$ that determines the running of couplings. 
The reason is simple to understand: Asymptotic safety respects Lorentz invariance in its construction and the only Lorentz-invariant scale that pertains to a free particle 
is its mass. Thus a particle with mass far below the Planck scale can never 
probe the non-trivial quantum gravity regime.
This is different in the case of a scattering event, as the one considered here, as the four-momentum of the intermediate particle, which does not fulfil an on-shell condition, 
can probe high momenta if the center-of mass energy of the scattering is high enough. Thus we set $G_N= G_N(s)$ in this part of our calculation.

\begin{figure}[!here]
\begin{center}
\includegraphics[width=0.7\linewidth]{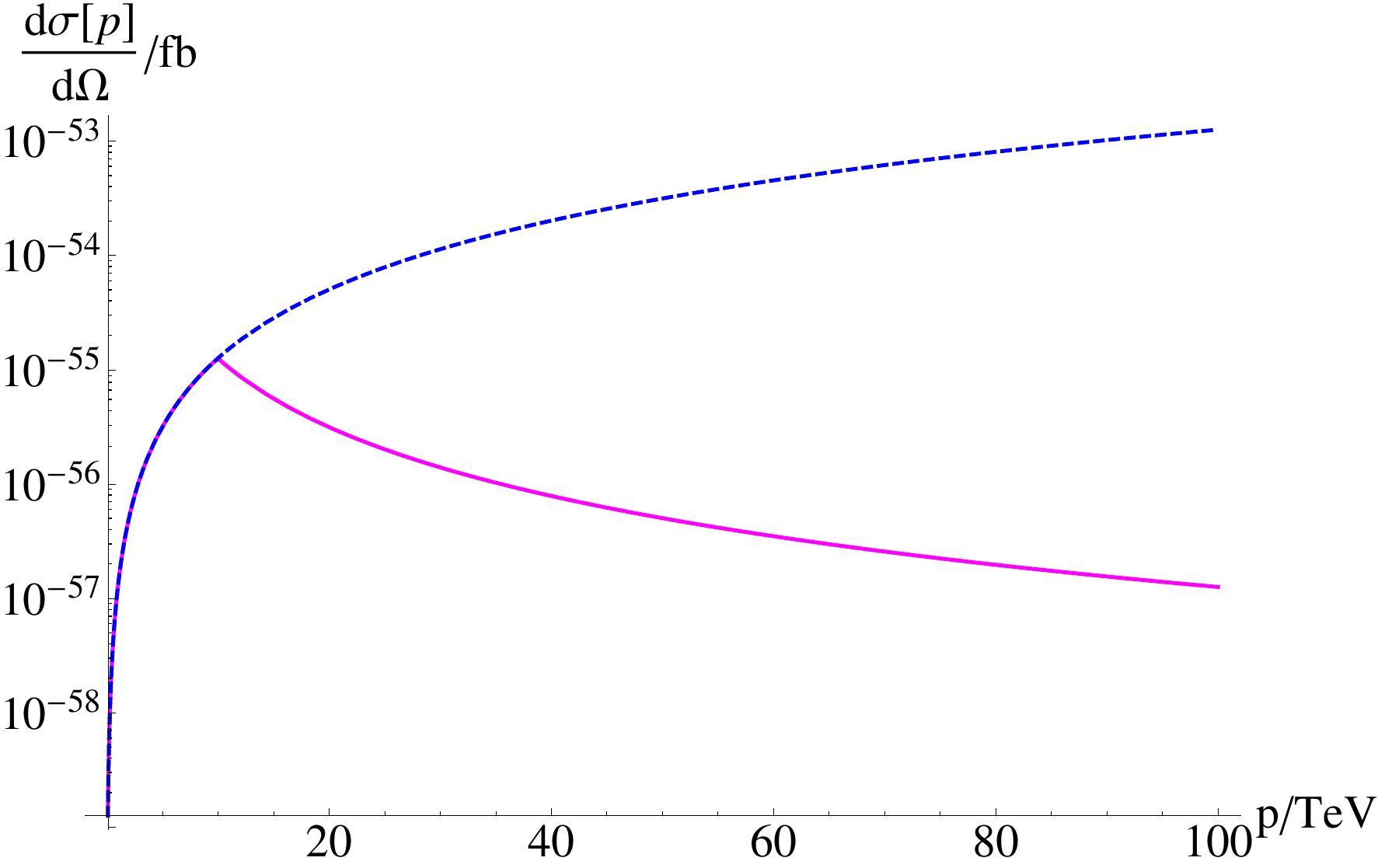}
\end{center}
\caption{We depict the cross section at tree-level for $\gamma- \gamma$ scattering via a graviton at a scattering angle of $\frac{\pi}{2}$ in the case of $G_N= \rm const$ 
(blue dashed line) and the asymptotically safe case (magenta line). The transition scale, which we have chosen to be $k_{\rm trans}=10\,  \rm TeV$ for the purpose of illustration, 
is clearly visible, since the asymptotically safe cross section decreases for $p^2>k_{\rm trans}^2$. Note that with a smoother crossover between the classical, 
constant and the quantum, scaling behaviour of Newton's coupling, the sharp transition in the cross section would be removed.
The Standard Model contribution for  $\gamma- \gamma$- scattering lies many orders of magnitude above the cross-section via graviton exchange
and is not included here.  \label{Guptaplot}}
 \end{figure}

Implementing the scale dependence \Eqref{runningG} in the cross section \Eqref{Gupta} in $d=4$ allows for a comparison to the case of a constant Newton coupling, see 
fig.~\ref{Guptaplot}. For the purpose of illustration we have chosen a low transition scale $k_{\rm trans}= 10\, \rm TeV$ here, but let us emphasise again that this scale is yet 
to be determined from observation.

As is well known, this cross section is much too small to be measurable at energy scales which are accessible experimentally today, and also suppressed by several orders of 
magnitude in comparison to the Standard Model contribution at these energies (see fig. \ref{gg_SM_AS} for the Standard Model contribution).

The crucial feature that we present here, is that within asymptotic safety, it does not suffer from a unitarity problem, as it does in a perturbative treatment. The fixed-point 
behaviour of the dimensionful Newton coupling, see \Eqref{runningNewton} causes the cross section to be finite at arbitrarily high energies, and in particular 
induces a decay $\sigma(p) \sim \frac{1}{p^2}$ at high energies, see also fig. \ref{Guptaplot}. 
This ensures perturbative unitarity, as a necessary condition is that the 
cross section is bounded by $\frac{1}{s}$.  \\

\section{Measuring photon-photon scattering in the laboratory}\label{gammagammalab}

In the following, we will analyse the same process in a scenario with $n= d-4$ large extra dimensions 
\cite{ArkaniHamed:1998rs,Antoniadis:1998ig}, where we assume that the extra dimensions form an $n$- torus.
 Let us explain our interest in extra dimensions in view of the current experimental findings at the LHC, which have started to exclude the existence of extra dimensions 
with a Planck scale at a few TeV \cite{Chatrchyan:2011fq, ATLAS:2011ab}.

A major motivation for extra dimensions comes from particle physics: Large extra dimensions have mainly been proposed 
to solve the hierarchy problem in the Standard Model. To this end, they should be accessible experimentally at the LHC. A fundamental Planck scale $M_{\ast}$ above the 
electroweak scale decreases the degree of fine-tuning in the Higgs sector, but only removes it completely if $M_{\ast}$ agrees with the electroweak scale.
However, the experimental falsification of such a solution to the hierarchy problem in its simple form by no means implies that a scenario with extra dimensions is not interesting.
Apart from the motivation from particle physics, there is another reason why extra dimensions might 
exist. In many quantum gravity scenarios, spacetime emerges from some underlying fundamental structure, see, e.g. \cite{Dowker:2005tz, Baratin:2011aa}, and it can 
easily be imagined that our universe has more than (3+1) dimensions, and a non-trivial spacetime topology\footnote{A tentative connection of asymptotic safety to such 
scenarios can arise, e.g. by a second order phase transition from a quantum geometrical phase to a continuous phase, see also \cite{Percacci:2010af}.}. 

Phrased differently, the observation of a four-dimensional spacetime at classical scales only restricts the spacetime topology and the number of 
dimensions in the infrared regime, where quantum gravity effects are negligible. 
Non-trivial quantum gravity dynamics can then allow for the existence of extra dimensions on smaller scales. 
Consider, e.g. a sum-over-histories, i.e. path-integral, approach for quantum gravity, where in a most general setting spacetime dimensionality and 
topology should not be fixed by hand, but be emergent in the sense that the expectation value for the dimensionality 
should be four at large scales, and spacetime should be an approximate Minkowski spacetime on large scales. In this setting, 
it is easily conceivable that the spacetime topology is much more complicated on smaller scales, and the existence of large extra dimensions could be one possible manifestation.

So, quite apart from the 
particle-physics motivation for extra dimensions, these can exist due to dynamics of quantum gravity, and then need not be connected 
at all to the electroweak scale. 
Thus analysing a higher-dimensional case is of general interest.

Within the asymptotic-safety scenario, it is not necessary to postulate the existence of extra dimensions for consistency reasons. 
Nevertheless, evidence so far suggests that the interacting 
fixed point underlying the scenario exists in $d \geq 4$ \cite{Fischer:2006fz}. Thus asymptotic safety is well-compatible with the existence 
of extra dimensions and a non-trivial spatial topology. 

From an observational point of view extra dimensions are obviously interesting as they bring 
down the momentum scale where quantum gravity effects become observable, since
$M_{\ast}^{n+2}=\frac{ M_{\rm Planck}^2} {(2 \pi r)^n} $ is the fundamental Planck scale, which can be considerably lower than $M_{\rm Planck}$, depending on the 
radius $r$ of the extra dimensions.
In the context of asymptotic safety and the existence of extra dimensions 
this has lead to predictions for different LHC processes \cite{Litim:2007iu,Gerwick:2011jw,Gerwick:2010kq}.

In the following section we want to investigate the implications of photon-photon scattering in this setting for three viable future experimental setups.
As pointed out,
 photon-photon scattering is particularly apt for beyond Standard Model searches due to the absence of a tree-level process. This has already lead to a number of experiments:
In the experimentally more feasible situation of photon propagation in external fields, the absence of a Standard Model ``background''
is already widely exploited in the search for beyond-Standard Model physics, see, e.g. \cite{Raffelt:1987im,Jaeckel:2010ni,Redondo:2010dp,Dobrich:2012sw}.

\subsection{Experimental options}
The natural prerequisite for such an undertaking is the availability of high-energetic photons with well-adjustable spectrum and direction.

\subsubsection{Linear electron collider}
Firstly, we study the option of photon-photon scattering at a possible ``International Linear Collider'' 
(ILC), the realisation of which will provide an advantageous environment for in-depth studies of the Standard Model and its proposed 
extensions \cite{DeRoeck:2009id}.
Such an electron accelerator facility can  also be used as a high-energy photon-photon collider by means 
of Compton back-scattering through the employment of a suitable laser source \cite{Ginzburg:1983}\footnote{Note that similar techniques
are already successfully employed at smaller electron linacs, see, e.g.  \cite{Albert:2010}.}. 
The photon-collider mode then opens up the possiblity to study several important and distinct particle physics processes in great detail \cite{Boos:2000ki}.
Note that under the employment of a UV-cutoff, it has been pointed out 
in \cite{Cheung:1999ja,Davoudiasl:1999di} that a future linear electron collider in a photon collider mode can be used to access the 
scales at which Kaluza-Klein graviton exchange becomes important if extra dimensions are realised in nature. 
Intriguingly, with an appropriate choice of the laser wave-length, the back-scattered photons can typically gain 
about 80\% of the beam energy of the electrons (for the maximal energy see fig.~\ref{photonenergy}).

Recent studies of an ILC-based photon collider with optimistic design
suggest that the corresponding luminosity will reach $\mathcal{O}(10)\, \mathrm{fb}^{-1}$ per year, see \cite{Bechtel:2006}. 
Center-of-mass electron energies of $E_{\mathrm{cm}}=2 \rm\, TeV$
then allow to reach photon energies of $0.8\, \rm TeV$ at similar luminosities. In a more conservative estimate with $E_{\mathrm{cm}} = 1 \, \rm TeV$,
we have photon energies of $0.4 \, \rm TeV$ at our disposal.

\subsubsection{Muon collider}

A muon collider, first proposed by \cite{Budker:1969cd}, see also \cite{Ankenbrandt:2007zz,Geer:2010zz}, presents an interesting option for a future collider, since the energy loss through synchrotron radiation is reduced considerably in comparison to the electron case, thus allowing to use ring accelerators instead of linear accelerators. The muon source is usually taken to be pions, which are generated in a collision of high-energy protons with a mercury target. Current technology could allow to accelerate muons up to several TeV within their lifetime.

\begin{figure}[!here]
\begin{center}
\includegraphics[width=0.6\linewidth]{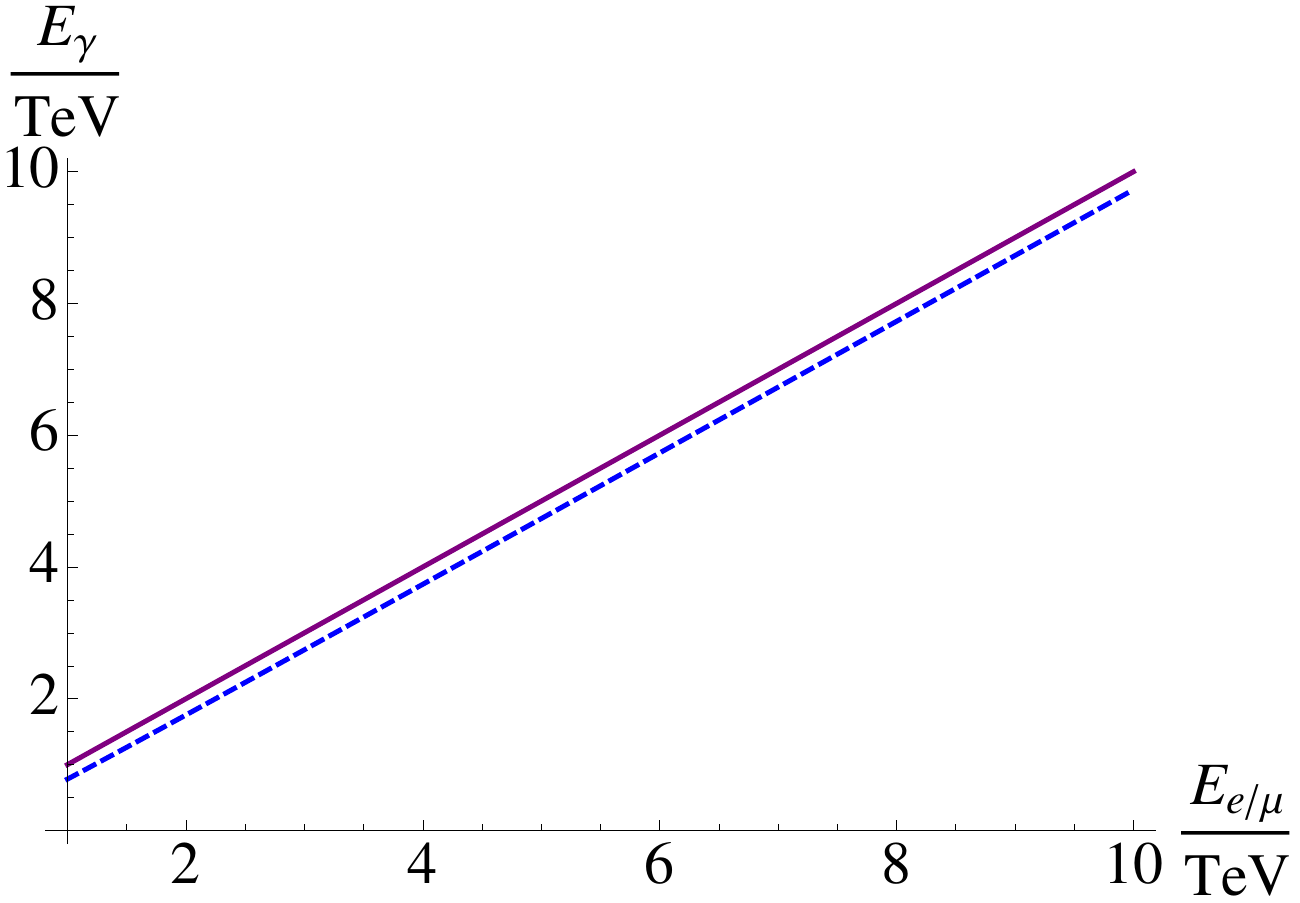}
\end{center}
\caption{\label{photonenergy} Here we plot the final photon energy as a function of the beam energy of an electron/ muon beam, respectively with initial photon energy $\omega_i= 10 \, \rm keV$. The purple line is for the case of an electron beam, whereas the dashed blue line is for the case of a muon beam.}
\end{figure}

High-energy photons can then again be produced by Compton backscattering the muons of a high-intensity laser. This setup is identical with the electron-accelerator case. From kinematics, it is straightforward to deduce the maximal photon energy in a setting with particle beam energy $E_{\rm beam}$ and final photon frequency $\omega_f$ (in the limit where for the initial photon frequency $\omega_i \ll E_{\rm beam}, \omega_{f}$  and also $m_{e/\, \mu} \ll E_{\rm beam}$):
\begin{equation}
\omega_f= 4 E_{\rm beam}^2 \omega_i \frac{1}{m_{e/\, \mu}^2+ 4\,E_{\rm beam} \omega_i}
\end{equation}
Clearly the maximal photon energy is slightly higher in the case of electrons, however it might be more feasible to reach higher beam energies for the muon case.
Using initial photons with energies of $\omega_i= 10 \, \rm keV$, final photon energies in the TeV range become accessible, cf. fig.~\ref{photonenergy}.
Initial photon energies in the $\mathcal{O}$(keV) range could be available from a synchrotron, free electron laser or even a small-area, purely laser-based source, cf. footnote \ref{fn:betatron}.

\subsubsection{Purely laser-based experiments}

Thirdly, let us note that there exists another viable option to produce very high-energetic
photons which does not rely on the employment of
a large accelerator facility. It comes with the rapid advancement in the
technology of high-intensity lasers\footnote{See, e.g. upcoming next-generation laser facilities such as the ``Extreme light infrastructure'' and the ``International Zeta-Exawatt Science Technology' \cite{eli_izest}.}:
Since the introduction of ``chirped pulse amplification'' \cite{CPA}, 
the achievable laser intensity has gone up by about seven orders of magnitude.
With current lasers exceeding intensities of $10^{22}\frac{\mathrm{W}}{\mathrm{cm^2}}$ and the implicit feature of 
strongly inhomogeneous fields,
tests of the QED strong field regime, including medium-like properties of the vacuum \cite{Gies:2011he,Karbstein:2011ja}, the Schwinger effect  \cite{Schutzhold:2008pz,Dunne:2009gi}
and the discovery of beyond Standard-Model physics \cite{Tommasini:2010fb,Dobrich:2010hi,Dobrich:2010ie} become accessible,
see, e.g. \cite{Tajima:1900zz,DiPiazza:2011tq}
for general overviews and opportunities.

The feature which is most relevant to our study is that high-intensity lasers in the optical regime
can be employed as electron accelerators and high-energy photon sources themselves.
In consequence it is possible to run a purely laser-based setup which creates high-energy photons
by Compton back-scattering\footnote{Currently also high-energetic photons with spectra peaking at up to $\mathcal{O}(100)$ keV  are also
directly available through betatron oscillation of the electrons in a laser-irradiated plasma \cite{plasma_betatron}.
Possibly, also this technique will can be upscaled with respect to the photon energy in future setups. \label{fn:betatron}}
 off laser-accelerated electrons  \cite{schwoerer}.
To allow for temporal synchronisation
of the electrons and the photons prior to the Compton back-scattering process, 
the laser beam first runs through a
beam splitter. One part of the beam is used for the electron acceleration, whilst the second part of the beam is focussed onto 
the accelerated electrons, where again high-energetic photons are 
produced in a back-scattering process.  The electron acceleration process can roughly be understood as follows:
A high intensity laser beam, if properly focussed
into a gas jet, can accelerate electrons from the jet, e.g. 
through the wake-field technique \cite{Tajima:1979bn}: It uses that electrons can ``surf" a plasma wave\footnote{The terminology arises as the electron can be regarded as ``surfing'' down the plasma wave being accelerated by the electric field within the plasma not unlike a surfer who gains kinetic energy.} which is excited inside a
medium through ponderomotive forces of the high-intensity beam: In a simplified picture, the plasma wave can be imagined to be generated by
the radiation pressure of the dense photon flux. Through breaking the plasma wave in a controlled manner, electrons can be injected into that wave which provides a co-moving, longitudinal electric field which can exceed the field available in conventional accelerators by far (in the plasma, one has, in present-day setups up to $\mathcal{O}(100)$GV/m).
 Of course, with the wake-field acceleration technique, a challenge lies in maximizing the distance over which the electron is accelerated which necessitates to guide the laser pulse within the plasma. 
At present, electrons can reach energies at the GeV-scale with this technique \cite{Leemans:2006}, with acceleration
lengths around 3cm in the cited experiment.
However, in principle also a multi-stage acceleration is conceivable: 
Fast electrons could be created via the wake-field technique in a first stage, while in consecutive stages,
which must be temporally well synchronised, the electrons could be further accelerated within graded plasma wakes.
In such a setup, the electron energy scales linearly with the number of stages.
In addition, a reduction of the electron density in the plasma could increase the available electron energies
by several orders of magnitude. By employing both these lever arms, in ambitious setups
the TeV scale is accessible \cite{Tajima:2011zza,Tajima:2010yt}.

With photon energies up to $\sim 0.8\, \rm TeV$ available in an ILC-type setup, and possibly even higher
in a muon collider or a fully laser-based experiment, let us now analyse photon-photon scattering 
at these energies within a scenario involving asymptotic 
safety and extra dimensions, to search for observational footprints of asymptotic safety.\newline\\

\subsection{Set-up of the calculation}
The crucial point in our calculation is again the incorporation of non-trivial quantum gravitational effects through a running Newton coupling. The coupling has dimensionality $2-d$ 
in $d$ spacetime dimensions, thus the generalisation of the scaling behaviour \Eqref{runningNewton} reads:
\begin{equation}
G_N(k) = \frac{G_{\ast}}{k^{d-2}}= \frac{G_{\ast}}{k^{n+2}},\label{runningG_n}
\end{equation}
for $d= 4+n$.
Here it is suitable to translate the running Newton coupling into a wave-function renormalisation $Z(k)$ for the graviton by using $G_N(k)= G_{\rm Newton} Z^{-1}(k)$, 
where $G_{\rm Newton}$ is the constant Newton coupling that we measure at low momentum scales, and the scale dependence is now carried by  the wave-function renormalisation. 
The graviton propagator 
is now schematically (neglecting the tensor structure) given by 
\begin{equation}
D(p)= \frac{1}{Z(k)\, p^2}.
\end{equation}

In the standard approach to extra-dimensions gravity, with a non-running $G_N$, the summation over the tower of Kaluza-Klein (KK) gravitons in diagrams such as those in fig.~\ref{diags} leads to an ultraviolet divergence for $n \geq 2$. Thus one usually 
resorts to the introduction of an explicit cutoff-scale $\Lambda$ to regularise the sum. This procedure has been followed in \cite{Cheung:1999ja,Davoudiasl:1999di,Atwood:1999cy}. 
Here we will 
compare results obtained in this way to 
results that arise when asymptotic safety is postulated as the UV completion for gravity.

Thus, we perform the sum over the Kaluza-Klein modes in the following way: As the mass-splitting between two consecutive Kaluza-Klein modes is very small compared to our
 center-of mass energy\footnote{As an example, the mass-splitting for $M_{\ast}= 1 \,\rm TeV$ and $n=2$ 
is $\delta_m \sim 3 \cdot 10^{-3} \rm eV$.}, we rewrite the sum into an integral, see also \Eqref{KKmass1}. 
We identify $m^2$ as the relevant RG scale (for a discussion of other choices see \cite{Gerwick:2011jw}), since in our setup $s \ll M_{\ast}^2$. 
Accordingly low-lying KK states probe 
the classical regime, whereas KK states with high masses are subject to fixed-point behaviour. For setups with $s \simeq M_{\ast}^2$, 
the identification of the RG scale should obviously also include $s$.

The integral to be performed at the amplitude level, see \cite{Giudice:1998ck} then reads
\begin{eqnarray}
&{}&\int_0^{\infty} dm \frac{m^{n-1}}{Z(m)\left(s-m^2\right)}= \int_0^{k_{\rm trans}}dm \frac{m^{n-1}}{s-m^2}+ k_{\rm trans}^{n+2}\int_{k_{\rm trans}}^{\infty} dm \frac{m^{n-1}}{m^{n+2}\left(s-m^2\right)}.\label{KK_ints}
\end{eqnarray}
and similarly in the $t$ and $u$ channel. Here, the first term is identical to the full contribution from the effective theory with a cutoff at $k_{\rm trans}$, where 
the second term is new, and gives the contribution of Kaluza-Klein masses in the fixed-point regime. The factor $k_{\rm trans}^{n+2}$ arises from the dimensionality of the 
Newton coupling, see \Eqref{runningG_n}.

In the $s$ channel, we find a pole when the Kaluza-Klein graviton is on-shell; we use the principal value prescription to evaluate the integral. Since for our 
experimental setting we are interested in $s < k_{\rm trans}^2$, the pole always occurs in the ``classical" part of the integration. For details on this calculation, see appendix
\ref{sec:app_mass}.

Let us note that our results differ from those in \cite{Cheung:1999ja,Davoudiasl:1999di,Atwood:1999cy}, where the summation over KK states in the $s$, $t$ and $u$ 
channel was set to an equal result.
As can be seen easily, this removes the pole at scattering angle $\theta=0$, which already occurs in the case without extra dimensions, cf. \Eqref{Gupta}, 
since it naturally arises from the $t$ and $u$ channels. It arises as for $\theta=0$ 
the massless graviton, i.e. the lowest state of the Kaluza-Klein tower, goes on-shell. The approximation in \cite{Cheung:1999ja,Davoudiasl:1999di,Atwood:1999cy} 
removes this pole.
To access the full dependence of the cross section on the scattering angle it is thus mandatory to avoid any such approximation.
Note however that for a viable experiment it is necessary to consider scattering angles $\theta>0$.

In the upcoming section, we compare our results with the Standard Model cross section. Note that the Standard-Model contribution at energies above the W-boson mass is dominated by a W loop, in comparison to which the fermionic loop 
can be neglected \cite{Jikia:1993tc}.
It is the unpolarised cross section exactly at these energies \cite{Gounaris:1998qk} which we are interested in.

\subsection{Results}
To discuss our results, we plot the integrated cross-section $\sigma$, integrated for the scattering angle $\theta$ in $\vert \cos (\theta) \vert < \cos (30^{\circ})$. This choice of integration region excludes the pole at $\theta =0, \pi$.

\begin{figure}[!here]
\begin{center}
\includegraphics[width=0.7\linewidth]{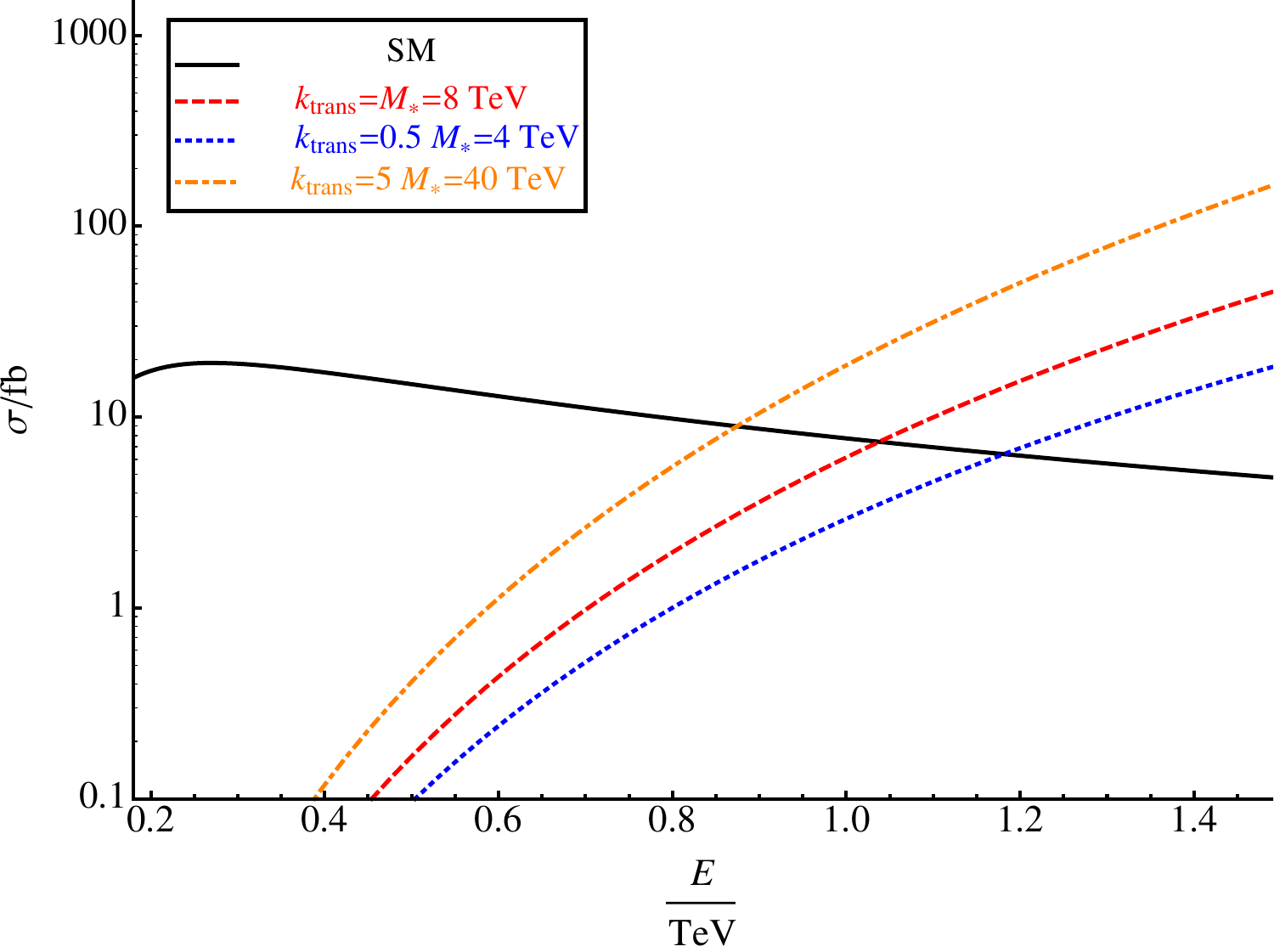}\\
\includegraphics[width=0.7\linewidth]{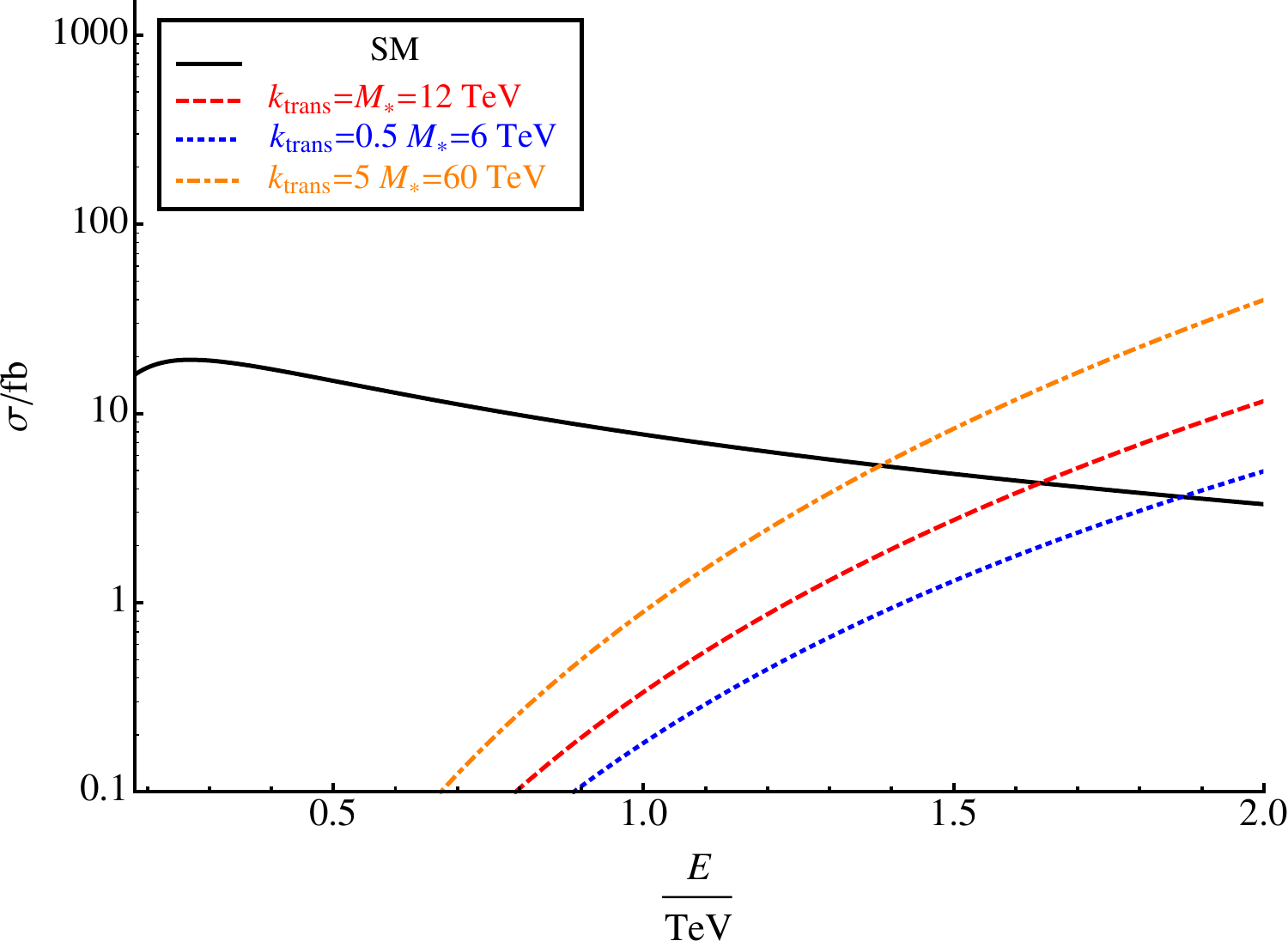}
\end{center}
\caption{\label{gg_SM_AS} Here we plot the integrated cross-section $\sigma$, integrated for the scattering angle  ($\pi/6\leq\theta\leq5 \pi/6 $) for the Standard Model (black thick line) and for asymptotic safety with $n=2$, 
where we show the following choices of transition scale and Planck scale: Upper panel: ($M_\ast=8 \,\rm TeV$, $k_{\rm trans}= 8 \,\rm TeV$) (red dashed), 
($M_\ast=8\, \rm TeV$, $k_{\rm trans}= 4 \,\rm TeV$) (blue dotted), ($M_\ast=8 \,\rm TeV$, $k_{\rm trans}= 40\, \rm TeV$) (orange dot-dashed); lower panel: ($M_\ast=12 \,\rm TeV$, $k_{\rm trans}= 12 \,\rm TeV$) (red dashed), 
($M_\ast=12\, \rm TeV$, $k_{\rm trans}= 6 \,\rm TeV$) (blue dotted), ($M_\ast=12 \,\rm TeV$, $k_{\rm trans}= 60\, \rm TeV$) (orange dot-dashed)}
\end{figure}

The first important conclusion is that the total cross section via graviton exchange becomes of the same order of magnitude as the Standard Model contribution 
\cite{Gounaris:1998qk} with extra dimensions at the TeV scale, see fig.~\ref{gg_SM_AS}.

Photon energies of about $0.8\, \rm TeV$ suffice to  clearly distinguish the pure Standard Model cross section from the cross section arising from 
asymptotic safety with $M_{\ast}=8\, \rm TeV$ and $n=2$ in this setting. We conclude that using available technology to construct a dedicated photon-scattering experiment 
allows to start constraining the parameter space for asymptotically safe quantum gravity. Experiments can access the quantities $(M_{\ast}, n , k_{\rm trans})$
 beyond the regime accessible at the LHC, and thus 
start to uncover the structure of spacetime, and find the fundamental degrees of freedom of quantum gravity.

Scenarios with a higher transition scale are more easily detected experimentally, due to the following simple reason: In the fixed-point 
regime, contributions of Kaluza-Klein gravitons are suppressed due to the fixed-point scaling $G(p)\sim \frac{1}{p^{d-2}}$. Thus, with a higher  
transition scale, more Kaluza-Klein gravitons contribute in an unsuppressed fashion to the total amplitude. Therefore the asymptotic-safety 
scenario with a low Planck scale, but a transition scale above this scale would be most favourable for experimental detection.\\
Note the following important point about such a construction: Placing the transition scale far above the Planck scale one might worry that black-hole production would dominate 
any scattering process at these energies. It is crucial to realise that within asymptotic safety, modifications to the effective equations of motion for gravity 
will generically arise at scales lower than 
the Planck scale. Thus entering a fixed-point regime far above the Planck scale by no means necessarily implies black-hole production in scattering experiments, 
see \cite{Basu:2010nf,Falls:2010he}.
Note however that our calculation, in which we have assumed that further operators in the effective action are not important for a leading-order evaluation of the 
cross section for photon-photon scattering, will presumably not be valid if $k_{\rm trans} \gg M_{\ast}$.

\subsubsection{Dependence on $n$}\label{ndepsec}

Let us now examine how the number of extra dimensions $n$ affects our result.

\begin{figure}[!here]
\begin{center}
\includegraphics[width=0.7\linewidth]{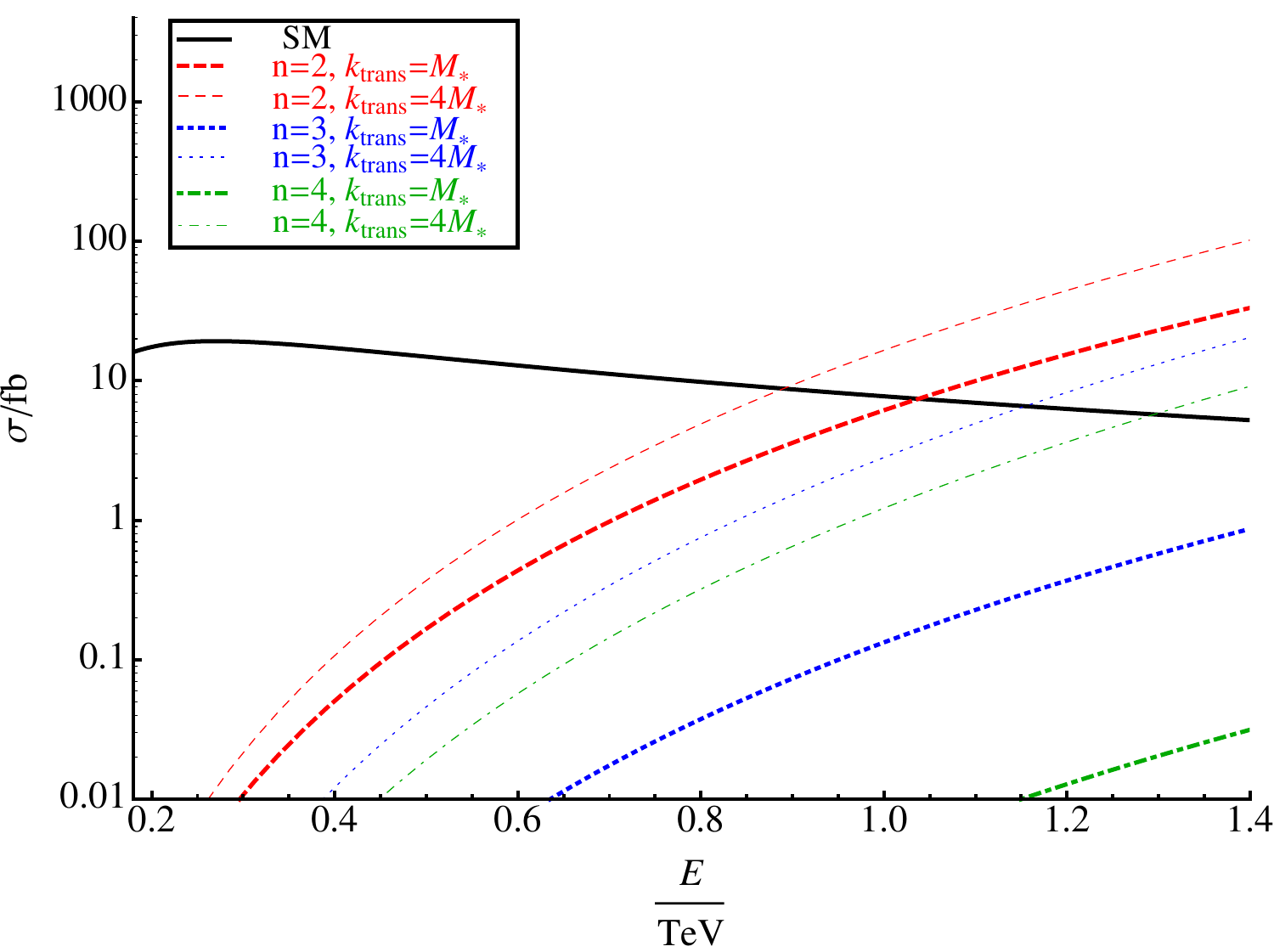}
\end{center}
\caption{\label{gg_SM_AS_na} Here we plot the integrated cross section ($\pi/6\leq\theta\leq5 \pi/6 $) for 
the Standard Model (black continuous line) and for asymptotic safety
with $M_{\ast}= k_{\rm trans}= 8\, \rm TeV$  (thick lines), and $k_{\rm trans}= 4 M_{\ast}= 32 \,\rm TeV$ where we 
show $n=2$ (red dashed lines), $n=3$ (blue dotted  lines) and $n=4$ (green dot-dashed lines). }
\end{figure}

For $k_{\rm trans}< M_{\ast}$, a higher number of extra dimensions at a fixed value of $M_{\ast}$ leads to a smaller signal, see fig.~\ref{gg_SM_AS_na}. 
This is due to the fact that the density of Kaluza-Klein states is $\sim m^{n-1}$. Accordingly, for higher $n$, less states will lie in the region $\{0, k_{\rm trans}\}$.
Thus less modes will contribute to the lower, ``classical" part of the integral 
in \Eqref{KK_ints}, and more modes will feel the suppression due to 
the fixed-point scaling of $G_N$. In this case the total cross section is lowered.
The effect that at $k_{\rm trans}< M_{\ast}$ 
scenarios with higher $n$ tend to elude experimental detection has also been observed in \cite{Davoudiasl:1999di, Cheung:1999ja}.

On the other hand, this effect becomes reversed for $k_{\rm trans} \gg M_{\ast}$:
Since the KK-integral in the fixed point regime comes with a factor of $k_{\rm trans}^{n+2}$, due to the dimensionality of the Newton coupling, a scenario 
with $k_{\rm trans}> M_{\ast}$ enhances the cross section more strongly for larger $n$. In this scenario it is thus easier to access settings with larger $n$.

\subsubsection{Asymptotic safety vs. momentum cutoff}

\begin{figure}[!here]
\begin{center}
\includegraphics[width=0.7\linewidth]{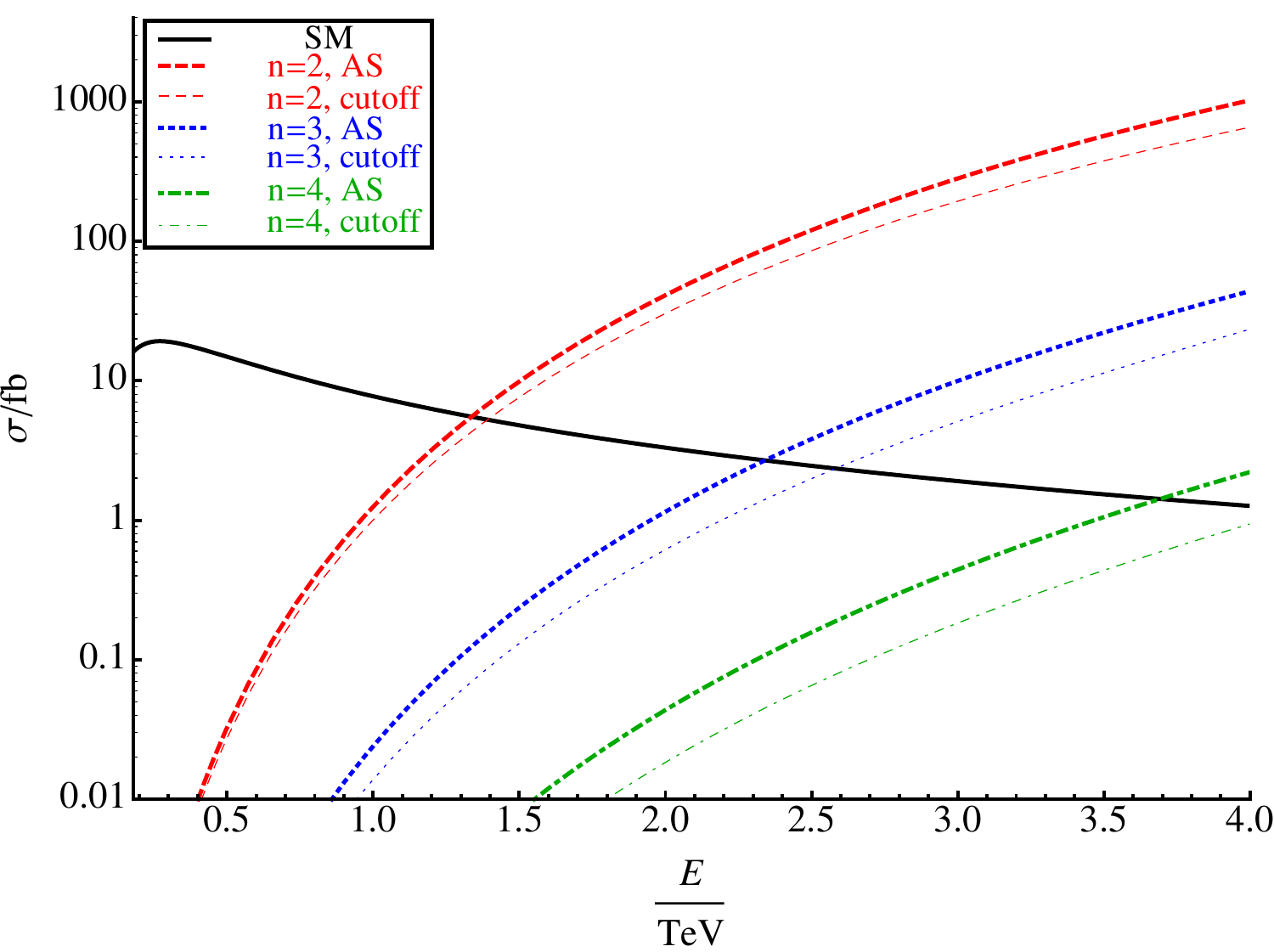}
\end{center}
\caption{\label{gg_SM_AS_n} Here we plot the integrated cross section ($\pi/6\leq\theta\leq5 \pi/6 $) for 
the Standard Model (black continuous line), for the cutoff theory (thin lines) with $M_{\ast}= 10\rm TeV$ and for asymptotic safety (thick lines) 
with $M_{\ast}= k_{\rm trans}= 10 \rm TeV$, where we 
show $n=2$ (red dashed lines), $n=3$ (blue dotted lines) and $n=4$ (green dot-dashed lines). }
 \end{figure}

One crucial question is: Is asymptotic safety in this setting distinguishable from other UV completions for gravity? To answer this question, we compare asymptotic 
safety to the result from a theory with a simple momentum cutoff.

As can be seen in fig.~\ref{gg_SM_AS_n}, the cross section in the asymptotically safe theory is larger than in the cutoff theory (Note that \cite{Cheung:1999ja,Davoudiasl:1999di} use a different convention for the fundamental Planck scale).
This is simply due to the fact, 
that the asymptotically safe cross section consists of two parts, namely the classical part, where the summation over Kaluza-Klein modes is identical to the result 
in the cutoff theory, and the Kaluza-Klein modes in the fixed point regime, see \Eqref{KK_ints}.\\
Clearly the difference between asymptotic safety and the cutoff theory becomes more pronounced with an increasing $n$. This is due to the fact that for larger $n$ 
the upper part of the integral over Kaluza Klein modes in \Eqref{KK_ints} starts to dominate over the lower part. The main reason again is that the density of Kaluza-Klein 
states grows as $m^{n-1}$, thus for larger $n$ a larger part of the modes falls into the fixed-point scaling regime.
Thus e.g. a  scenario with $n=4$ and $M_{\ast}=10 \,\rm TeV$ becomes more easily accessible at $E \sim 3\, \rm TeV$ within asymptotic safety, while more data and a higher experimental precision is needed within the cutoff theory, where the graviton cross section is nearly one order of magnitude below the asymptotic safety contribution.

 \begin{figure}[!here]
\begin{center}
\includegraphics[width=0.7\linewidth]{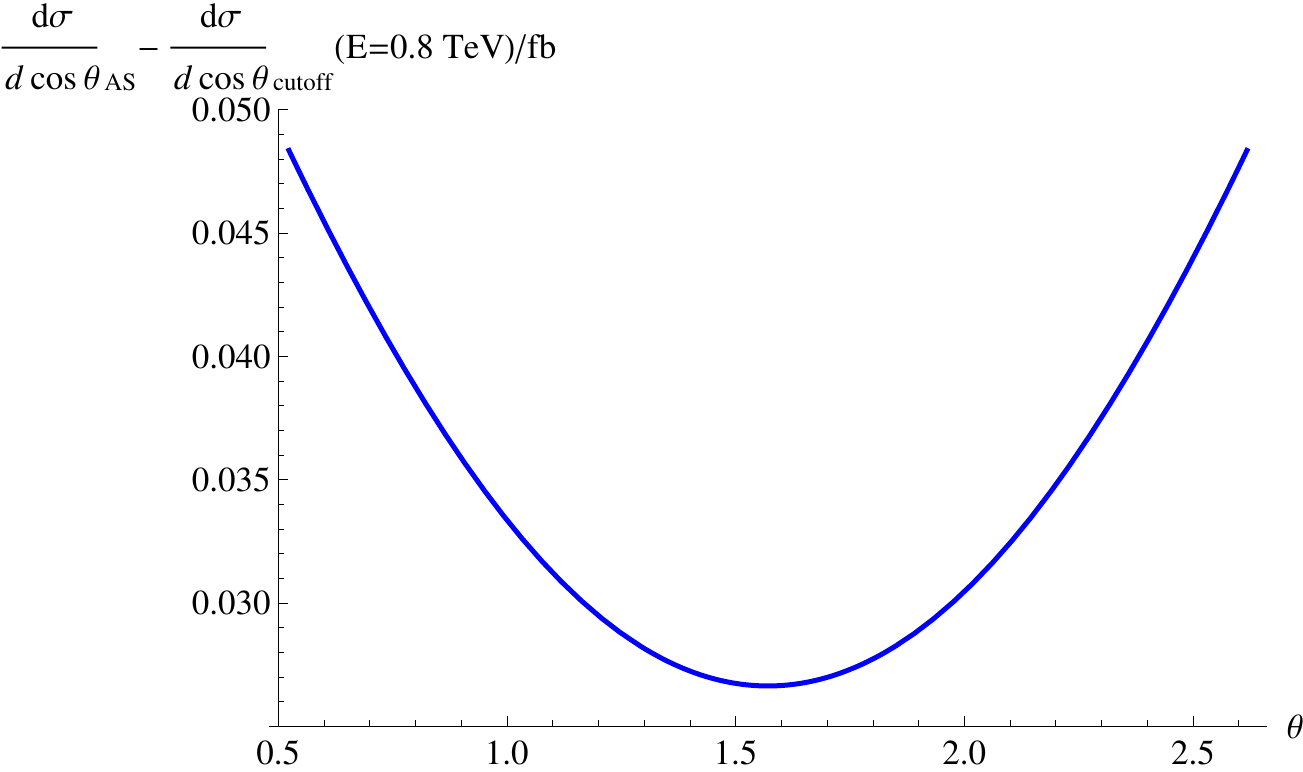}
\end{center}
\caption{\label{cutoff_AS_diff} We show the difference of the differential cross section $\frac{d \sigma}{d \cos \theta}$ in the theory with a 
simple UV cutoff at $M_{\ast}$, and asymptotic safety with $k_{\rm trans}= M_{\ast}$. We choose $E=0.8\, \rm TeV$ and 
$M_{\ast}=10\, \rm TeV$ and plot $n=2$.}
 \end{figure}

The unique signature of asymptotically safe quantum gravity is in principle most easily accessible with $s \sim k_{\rm trans}^2$, where $s$ starts to probe the fixed-point 
regime. The crucial difference to models with a cutoff that are usually considered is, that in these models new physics, i.e. new degrees of freedom, start to emerge 
beyond the cutoff scale. In asymptotic safety, the scale above which the cross section starts to decay is {\it not}
associated with the existence of any new degrees of freedom. 
The existing degrees of freedom, namely metric excitations and the Standard Model particles, show a dynamics which is governed by the fixed point, and leads to the decay of the 
cross section. No new degrees of freedom are necessary to obtain a theory that is finite and predictive in the ultraviolet.

At energies $s < M_{\ast}^2$ the signature of asymptotic safety can also be accessed in the angular dependence of the cross section, see fig.~\ref{cutoff_AS_diff}. Here we use the definition of the angular cross-section $\frac{d \sigma}{d \cos \theta} = 2 \pi \frac{1}{64 \pi^2 E^2} \vert \mathcal{M} \vert^2$, where the amplitude $\mathcal{M}$ is the sum over the diagrams in fig.~\ref{diags}, with a summation of Kaluza-Klein gravitons on the internal line. 

Let us stress that our result also holds in the case where all Standard-Model fields propagate into the extra dimensions, i.e. in a scenario with universal extra dimensions. In that case, there is a tower of W-boson Kaluza-Klein states, which will also contribute to the cross-section, and furthermore a part of the photons in such scattering experiments will show up as a missing-energy signal, when photons propagate into the extra dimensions. However for the case that we observe the scattering of two photons, the graviton-contribution is not changed. 

\subsubsection{Accessibility of $M_{\ast}$ \label{access_Mstar}}

To determine which range for possible values for the fundamental Planck scale is accessible within this setting, we define the quantity
\begin{equation}
\sigma_{\rm diff}= \frac{\frac{{\rm d} \sigma_{{\rm AS}}}{{\rm d} \cos \theta}-\frac{{\rm d} \sigma_{{\rm SM}}}{{\rm d} \cos \theta} }{\frac{{\rm d} \sigma_{{\rm SM}}}{{\rm d} \cos \theta}},
\label{eq:sigma_diff}
\end{equation}
which is a simple measure of the visibility of the asymptotic-safety (AS) signal over the Standard Model (SM) background. Depending on the luminosity and 
energy resolution of the experimental setup, different fundamental Planck scales can be accessed. Here we assume that a 
1 \% effect in the total cross section (i.e. Standard Model as well as graviton contributions) suffices for a detection.

 \begin{figure}[!here]
\begin{center}
\includegraphics[width=0.7\linewidth]{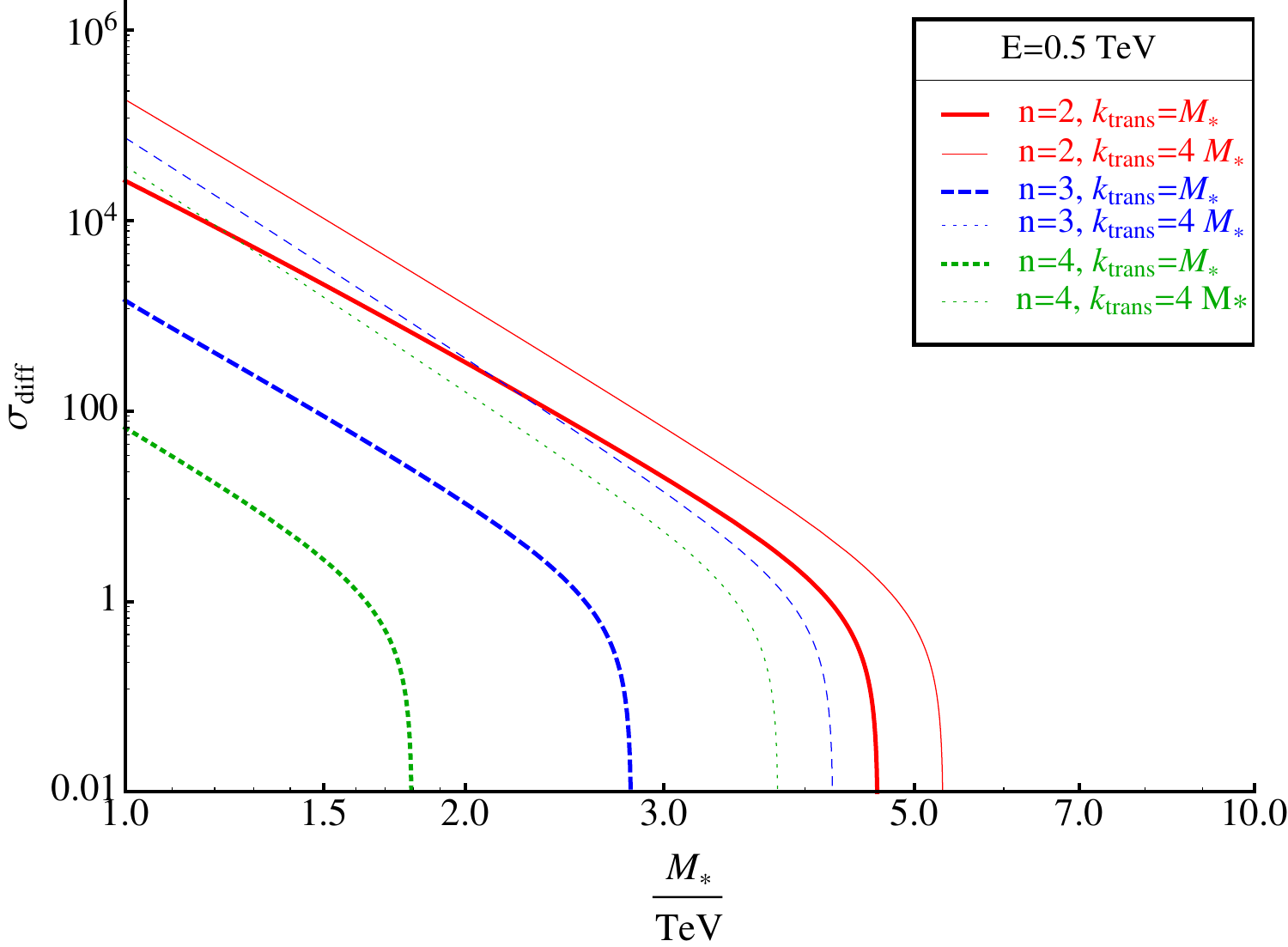}\\
\includegraphics[width=0.7\linewidth]{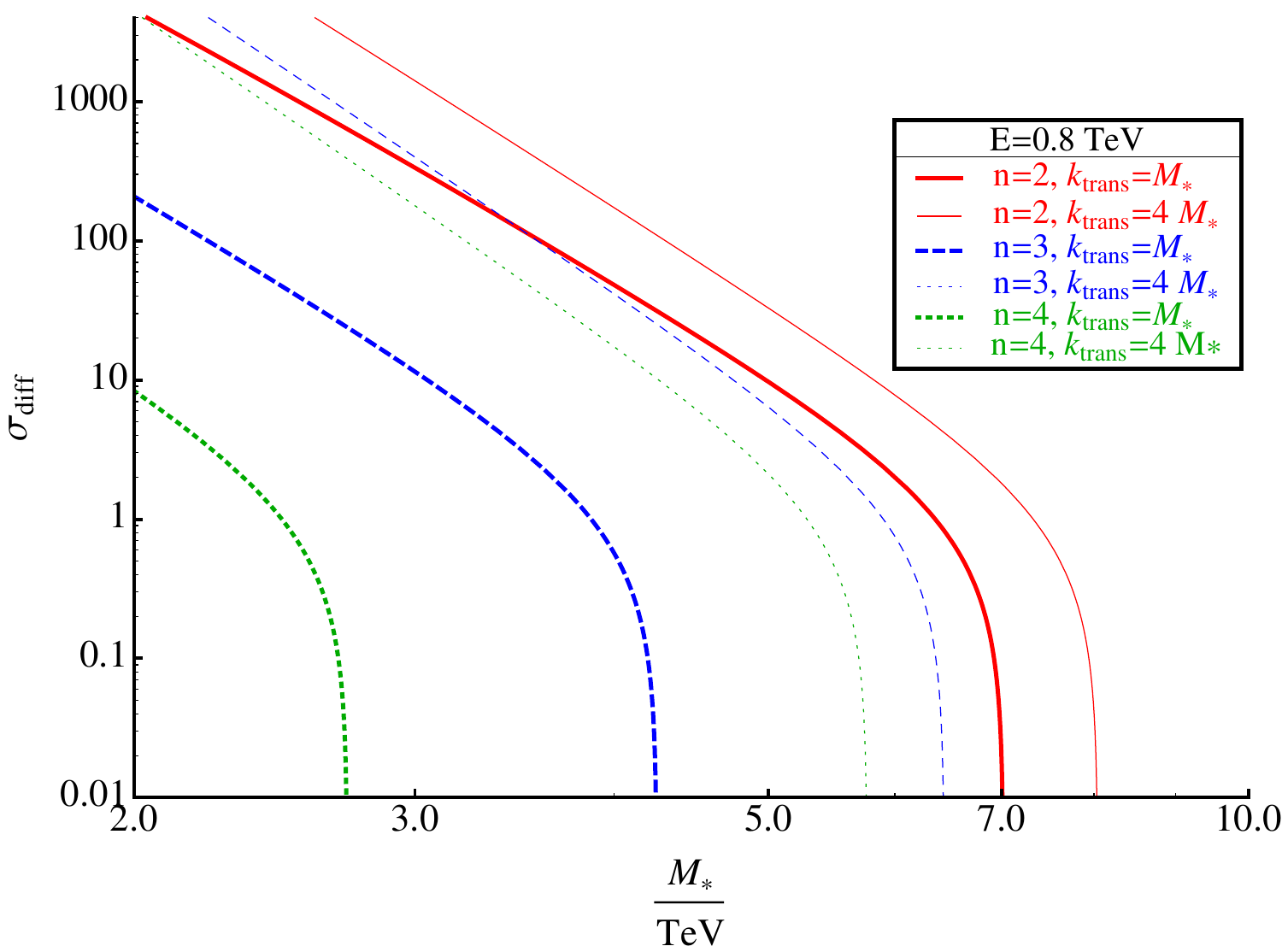}
\end{center}
\caption{\label{sigmadiff} Here we depict the quantity $\sigma_{\rm diff}$, cf. \Eqref{eq:sigma_diff}, as a function of the
 fundamental Planck scale at fixed scattering angle $\theta= \pi/2$ and fixed photon 
energy $0.5\, \rm TeV$ (upper plot) and $0.8\, \rm TeV$ (lower plot).
We show $k_{\rm trans}= M_{\ast}$ with thick lines, and $k_{\rm trans}= 4 \,M_{\ast}$ with thin lines. The colours 
denote different values of $n$ according to $n=2$ (red), $n=3$ (blue) and $n=4$ (green).}
 \end{figure}

Clearly a fundamental Planck scale of order $7\, \rm TeV$ is accessible with a photon energy of $E= 0.8 \,\rm TeV$, which could be reached at the ILC, cf. fig. \ref{sigmadiff}.

\begin{figure}[!here]
\begin{center}
\includegraphics[width=0.7\linewidth]{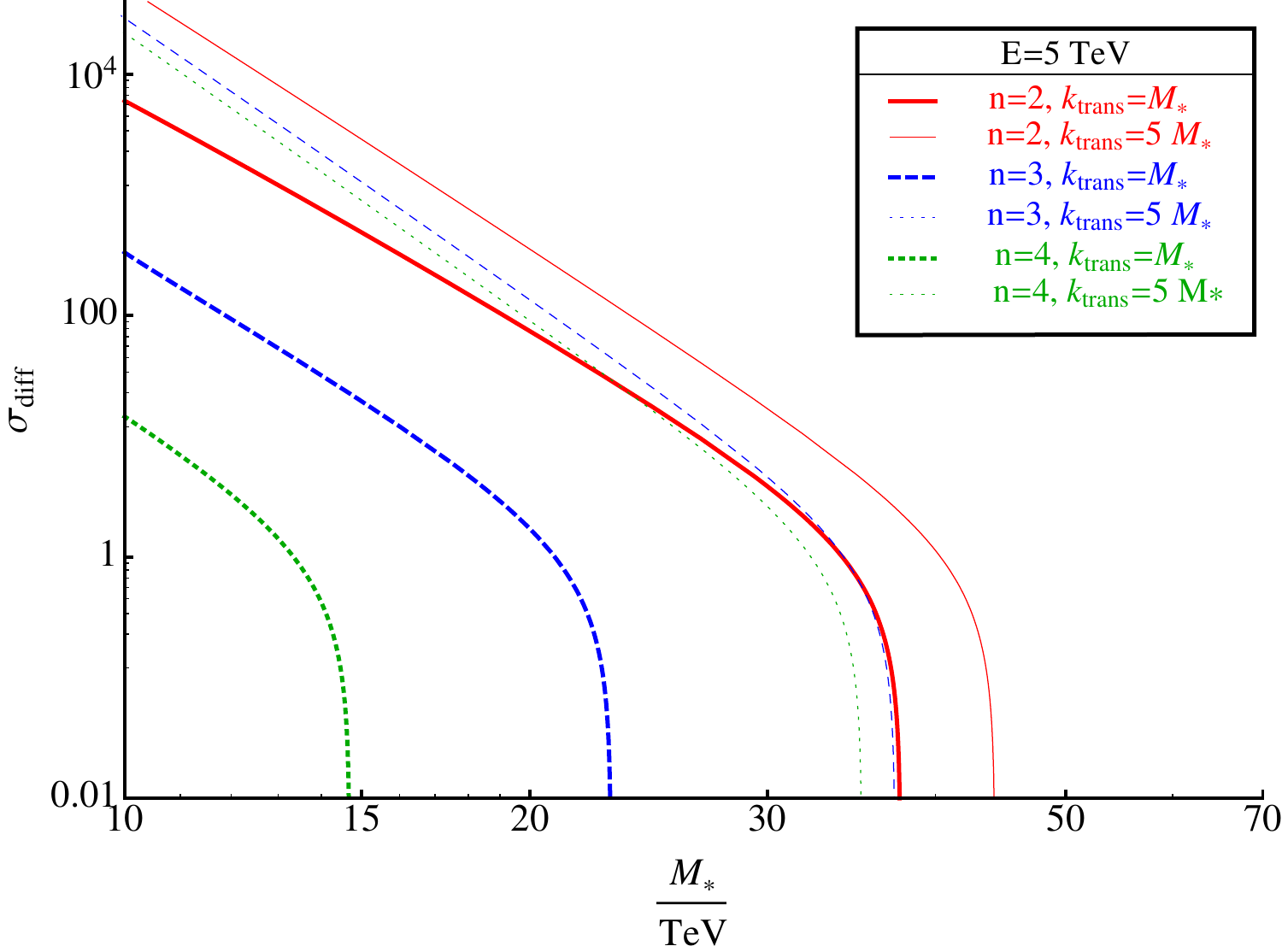}
\end{center}
\caption{\label{sigmadiffchall} Here we show the quantity $\sigma_{\rm diff}$, cf. \Eqref{eq:sigma_diff}, as a function of the
 fundamental Planck scale at fixed scattering angle $\theta= \pi/2$ and fixed photon 
energy $5 \rm TeV$ 
We show $k_{\rm trans}= M_{\ast}$ with thick lines, and $k_{\rm trans}= 5 M_{\ast}$ with thin lines. The colours 
denote different values of $n$ according to $n=2$ (red), $n=3$ (blue, dashed) and $n=4$ (green, dotted).
}
 \end{figure}

Let us add that in a very optimistic scenario, even higher electron energies might be used. A purely laser based setting using electron acceleration on a plasma wave excited by a high-intensity laser, and subsequent Compton backscattering may allow to reach photon energies of the order of $E_{\gamma} \sim \rm TeV$.
We point out that meeting this challenge is very worthwhile, since Planck scales of the order of $M_{\ast}=40\, \rm TeV$ become accessible in such a setup, see fig.~\ref{sigmadiffchall}.
Even though such a setup  pushes existing technologies to their very limit, 
it shows how precision measurements of processes without a Standard Model tree level contribution can probe the quantum gravity regime far above the 
energies that are actually necessary in the experiment. A photon energy of $5 \,\rm TeV$ suffices to access Planck scales of $M_{\ast}= 40 \,\rm TeV.$

Let us comment on the effect of further operators in the effective action on our result: Generically, the RG flow generates all operators which are compatible with the 
symmetries. For our process, only those contributing to a photon-photon-graviton vertex on a vanishing electromagnetic background (i.e. no electric or magnetic background fields), 
are of interest.
Similarly to the Euler-Heisenberg action, there will be an $F^4$ term induced by a graviton loop. This implies that it will be $\sim G_{N}^2$ and thus 
highly suppressed in the classical regime. 
In the fixed-point regime, this contribution can in principle become of the same size as the contribution from single-graviton exchange. As we have discussed, the contribution from the 
fixed-point regime is subleading, thus we expect a quantitative change in our results, which will however not alter our estimates considerably. The conclusions that photon energies of the 
order of several hundred GeV suffice to detect first experimental hints of quantum gravity should be robust.\\

Further operators that can contribute will be of the form $\rho_n \int d^dx \sqrt{g} \left(D^2 \right)^{2n}F^2$, where $\rho_n$ is the scale-dependent coupling of this term.
 Now there are two regimes in which we have to estimate the importance of 
this term, 
namely the classical regime, in which $G_N= \rm const$, and the fixed-point regime. 
In the former, we can assume that such couplings are constant, since they will be generated by gravitational fluctuations in the path integral, which can be neglected in the classical 
regime. Then a contribution to photon-photon scattering from such a vertex comes with additional factors of momenta. In the regime of interest, where $s <M_{\ast}^2$, this leads to a 
suppression of these terms in comparison with the term that we have accounted for.
Now let us comment on the fixed-point regime: For $n>0$ the coupling $\rho_n$ has negative dimensions of mass, thus it will scale as $\rho_n(p)\sim \rho_{\ast} p^{-4n}$ in the 
fixed-point regime. This scaling precisely compensates the additional factors of momenta in the vertices which this operator generates\footnote{If the vertices develop strong asymmetries in the external momenta, the scaling analysis becomes less straightforward, however any physical quantity must remain divergence free even in these more complicated cases.}. Thus this contribution to the full cross section 
has the same scaling properties as the term that we analysed, so the importance of such terms in the scaling regime in comparison to our result depends on the fixed-point values $\rho_{\ast}$. One may now assume that all matter couplings approach a Gau\ss{}ian fixed point at high energies, even when gravitational couplings approach an interacting one, see \cite{Harst:2011zx}. Thus $\rho_{\ast}=0$ in this scenario, and thus there is no contribution from higher-order operators in the fixed-point regime in this case.

\section{Conclusions and Outlook}
With several different approaches to a quantum theory of gravity available, which are based on different assumptions about the fundamental degrees of freedom, the 
realisation of symmetries, and the nature of spacetime itself, it is mandatory to find experimental evidence in order to understand what the fundamental 
properties of quantum gravity are.

Due to the weakness of the gravitational force at scales up to at least several $\rm TeV$, it is very difficult to access the quantum nature of gravity, and observable 
effects are expected to be very tiny. It is thus crucial to identify processes which carry a sizable quantum gravity signature.
Here, we study a Standard-Model scattering process without a tree-level contribution, namely photon-photon scattering. It arises purely from quantum effects, and is dominantly 
induced by an electron, resp. W-boson loop, depending on the energy scale.
In fact, the Standard-Model cross section for this 
process still awaits direct experimental confirmation, which will become available in the very near future  owing to the rapid progress in laser technology. \\
We examine the quantum gravity contribution to this process. Here, we make two central assumptions, namely that the UV completion for gravity is provided by asymptotic safety, and that extra 
dimensions exist.
The main property of asymptotically safe quantum gravity for this work is that the dimensionful Newton coupling runs in a power-like fashion in the fixed-point 
regime, cf. \Eqref{runningG_n}. This allows to calculate the cross section for the graviton contribution to photon-photon scattering at arbitrarily high energies, which respects 
perturbative unitarity at high energies due to asymptotic safety. \\
The assumption of extra dimensions is motivated not by particle-physics considerations, which requires the fundamental Planck scale to coincide with the electroweak scale 
in order for the hierarchy problem of the Standard Model to be solved. Instead we point out that a dynamics for quantum gravity which is based on a general sum-over-histories 
approach may allow for the existence of extra dimensions with a fundamental Planck scale at an energy scale, which need not be at all related to any particle physics scale.\\
Assuming the existence of extra dimensions is crucial to render the cross section due to graviton exchange of the same order of magnitude as the Standard Model cross section at 
experimentally accessible energy scales.
Let us stress, that a main advantage of our experimental setup lies in the fact that it uses a process which has no tree-level contribution from the Standard Model, but a tree-level 
contribution from quantum gravity. Processes of this type can have quantum gravity cross sections, which are of the same order as the Standard Model 
cross section, and which might very soon become measurable building on present-day technology.
In such processes, a tree-level quantum gravity contribution can become accessible in a setting with large extra dimensions, 
with a fundamental Planck scale of the order $M_{\ast}\sim \mathcal{O}(10 \rm TeV)$.
An optimistic setup, which builds on present-day technology, but most likely requires some work to get sufficient luminosity, even allows to access 
$M_{\ast}\sim \mathcal{O}(40 \rm TeV)$.

Indeed, in the first part of our study, we have first followed previous authors in the suggestion to test our scenario at the photon collider mode at the planned International
Linear Collider. We point out that photon energies of $0.8 \, \rm TeV$ suffice to detect signatures of asymptotically safe quantum gravity with a Planck scale up to 
$M_{\ast} \sim \mathcal{O}(10 \, \rm TeV)$. In particular, we discuss the possibility to distinguish asymptotic safety from a theory with a sharp momentum cutoff, by 
measuring the angular dependence of the cross section. We also show that within asymptotically safe quantum gravity, the cross section for photon-photon scattering 
is slightly higher than in the cutoff theory. Assuming a scenario where the transition to the fixed point regime occurs at a transition scale $k_{\rm trans}> M_{\ast}$, 
the cross section becomes enhanced significantly. 

Thus we conclude that a linear electron collider or a muon collider in the photon collider mode will allow to search for observational signatures of asymptotically safe quantum gravity.

In addition to this option, we have also studied a third viable option to produce high energetic photons, which could become available before the completion
of a future collider. With the rapid advance in high-intensity laser technology, a purely laser-based setup using electron wakefield acceleration might allow to reach photon energies at the $\rm TeV$ scale by exploiting the simple process of Compton backscattering. Such an experiment, which has the capability to test 
quantum effects within the Standard Model, as well as to search for many manifestations of new physics, provides an excellent possibility to test for quantum gravity effects. 
Within a setup that exploits currently available technologies to its very limits, Planck scales up to $40\,\rm TeV$ could become accessible.

Therefore the rapid advancement in laser technology might make a discovery of quantum gravity effects possible before these become accessible at collider based setups.

Let us emphasise that if confirmed, this candidate theory for quantum gravity offers a unifying picture of all Standard Model matter and gravity: 
Both are quantised as quantum field theories in the path-integral framework, using the existence of an interacting UV fixed point to ensure predictivity 
and finiteness of the theory. Establishing the possibility to quantise gravity in the well-tested framework of quantum field theory allows to straightforwardly 
study many of its properties, such as its imprints on scattering processes, particle oscillations etc. by means of standard quantum field theoretic tools. Our 
work is an example of how we can apply the framework of standard quantum field theory and in particular the effective action to deduce experimental results.

Let us comment on extensions of our work: A first is clearly to study the effect of higher order operators in quantitative detail. As we have argued, these
 should not affect our results at leading order, but quantitative precision requires their inclusion in further studies. In particular, it is crucial to investigate, 
whether the standard assumption that the matter degrees of freedom can approach a non-interacting fixed point in the UV, while gravity remains interacting, holds.
An extension of our work should also include the effects that fermionic matter has on our results. Since charged fermions couple to both photons and gravitons, the 
effective action will contain terms which contribute to the photon-photon scattering amplitude through diagrams containing internal fermions and gravitons. Within
 the one-loop effective action, results on the coupling between gravitons and photons induced by fermionic loops are already available, see, e.g. \cite{Davila:2009vt}, 
and could be used for extensions of our results.

A further interesting possibility arises, as photon-graviton mixing in a magnetic field is also possible within asymptotic safety. 
This effect has already been analysed in a scenario with extra dimensions and a momentum cutoff \cite{Deffayet:2000pr}. An extension to asymptotic safety is 
straightforward along the lines presented in this paper, and subject of ongoing work \cite{toappear}. Such a photon-graviton mixing can have 
exciting astrophysical consequences, related to the anomalous transparency of the universe to TeV photons. This might find an explanation in the 
following scenario: Mixing of photons with Kaluza-Klein gravitons implies that photons travelling to us over astrophysical distances, travel part of their 
way ``disguised" as a Kaluza-Klein graviton. This makes them immune to pair-production, which usually occurs when a TeV photon encounters a CMB photon or propagates within an, e.g. intergalactic magnetic background-field. 
A single Kaluza-Klein graviton has a negligible cross section for this effect to occur, thus a Kaluza-Klein graviton can traverse our universe without being 
lost to pair-production.

Finally let us add, that our work can also be extended to other UV completions for quantum gravity within the same framework: 
Although UV completions for gravity will differ on microscopic length scales, there presumably is a regime in which it is valid to use the
 metric as an effective degree of freedom, even though the microscopic degrees of freedom should differ. Thus, any UV completion for gravity can be 
studied within the effective-field-theory framework, where, starting from values for all the couplings determined from the microscopic theory, one can then 
follow the RG flow towards the infrared. These flows will then presumably differ for different UV completions over a certain range of scales, though they should 
agree in the infrared in order to correspond to our observable universe. Thus it is straightforward to compare predictions for processes such as photon-photon scattering 
from different underlying quantum gravity theories, by using the particular scale dependence of couplings that these induce. In principle, it is then possible to compare 
predictions from different quantum theories of gravity.

We conclude that photons provide for an interesting observational window into the quantum gravity domain, and might 
literally allow to make the effects of quantum gravity visible.

\acknowledgments
B.D. thanks the Perimeter Institute, where a large part of this work was completed, for its hospitality. Helpful discussions with J. Jaeckel, 
A. Ringwald and M.C. Kaluza are gratefully acknowledged. We thank H.~Gies and M.C. Kaluza for a careful reading of the manuscript.
B.D. acknowledges support by the DFG under grants SFB-TR18, GRK-1532 and Gi 328/3-2.\\
\\
Research at Perimeter Institute is supported by the Government of Canada through Industry Canada
and by the Province of Ontario through the Ministry of Research and Innovation.

\appendix
%
\section{Photon-Photon scattering with extra dimensions \label{sec:app_calc}}
%

For clarity we present some details of the calculation here.

We start from the action
\begin{equation}
 \Gamma=- \frac{1}{4} \int \mathrm{d}x \ \sqrt{-g}\, g^{\mu \kappa}g^{\nu \lambda} \ F_{\mu\nu} F_{\kappa \lambda} \ , \label{eq:action}
\end{equation}
with the electromagnetic field strength tensor $F_{\mu\nu}$ and the determinant of the metric tensor $g=\mathrm{det}(g_{\mu\nu})$.

From this, we obtain the two-photon graviton vertex on a flat background $\eta_{\alpha\beta}$, where we choose  $\mathrm{sign}(\eta)=(+1,-1,-1,-1)$.
We define the propagating gravitational degree of freedom $h_{\mu \nu}(x)$ via $g_{\mu \nu}(x)= \eta_{\mu \nu}+ h_{\mu \nu}(x)$.

We follow the index assignments as used in fig.~\ref{diags}. One then finds for the two-photon graviton vertex
\begin{align}
& V_{\mu \nu \alpha\beta}  = \label{eq:vertex}  -\frac{1}{2}\Big(p_{1}p_{2} \left( \eta_{\alpha \beta} \eta_{\mu\nu} - 
   \eta_{\mu\beta} \eta_{\nu\alpha} - 
   \eta_{\mu\alpha} \eta_{\nu\beta} \right) 
 - \eta_{\alpha\beta} p_{2,\mu} p_{1,\nu} + 
   \eta_{\nu\beta} p_{2,\mu} p_{1,\alpha} +  
  \eta_{\nu\alpha} p_{2,\mu} p_{1,\beta}\notag \\ &
  + \eta_{\mu\beta} p_{2,\alpha}  p_{1,\nu} + 
   \eta_{\mu\alpha} p_{2,\beta} p_{1,\nu} - 
 \eta_{\mu\nu} p_{2,\beta} p_{1,\alpha}  - 
   \eta_{\mu\nu}  p_{2,\alpha}  p_{1,\beta} \Big) \ ,
\end{align}
where in our convention, all momenta are chosen as incoming momenta.
Here, $\mu$ and $\nu$ label incoming photons with momenta $p_1$ and $p_2$, respectively, whereas $\alpha$ and $\beta$  denote graviton indices, cf. fig.~\ref{diags}.

We use the graviton propagator in harmonic gauge which
reads
\begin{equation}
 P_{k\, \alpha\beta\gamma\delta} (p)= \frac{16\pi G_N (k)}{p^2}\left( \eta_{\alpha\gamma}\eta_{\beta\delta}+\eta_{\alpha\delta}\eta_{\beta\gamma}-\eta_{\alpha\beta}\eta_{\gamma\delta}\right) \ .
\label{eq:prop}
\end{equation}

In a setting with large extra dimensions, the graviton propagator becomes
\begin{align}
& P_{k\,\alpha\beta\gamma\delta}(p)=  \frac{16\pi G_N(k)}{p^2-m^2}\Big[(\eta_{\alpha\gamma}\eta_{\beta\delta}+
\eta_{\alpha\delta}\eta_{\beta\gamma}-\eta_{\alpha\beta}\eta_{\gamma\delta})\notag
\\ &
-\frac{1}{2m^2}(\eta_{\alpha\gamma}p_{\beta} p_{\delta} + \eta_{\beta\delta} p_{\beta} 
p_{\gamma} + \eta_{\alpha\delta} p_{\beta} p_{\gamma}
 + \eta_{\beta\gamma}p_{\alpha} p_{\delta})
+ \frac{1}{6}\left(\eta_{\alpha\beta}+\frac{2}{m^2}p_{\alpha} p_{\beta}\right)
\notag
\\ &
\times \left(\eta_{\gamma\delta}+\frac{2}{m^2}p_{\gamma} p_{\delta}\right) \Big] \ ,
\label{eq:prop_kk}
\end{align}
see, e.g. \cite{Gerwick:2009zx}.
Note that, except for the global factor, all terms involving $m^2$ in \Eqref{eq:prop_kk} do not contribute on amplitude level for the tree-level-diagram under consideration due to transversality.

Taking into account the $d$-dimensional phase space for the final states,  the differential cross section is given by
\begin{equation}
\frac{d \sigma}{d \Omega} =\frac{1}{4}\frac{1}{8s}(2\pi )^{2-d}
\left(\frac{\sqrt{s}}{2}\right)^{d-4}\left(4\frac{t u }{s^2}\right)^{\frac{(d-4)}{2}} \vert \mathcal{M}\vert^2,
\end{equation}
where $\mathcal{M}$ follows from  Eqs. (\ref{eq:vertex}) and (\ref{eq:prop}).
As in the main text, Mandelstam variables $s = 4 E^2$, $t= -2E^2\left(1-\cos(\theta) \right)$ and $u = -2E^2\left( 1+\cos(\theta)\right)$ 
with the center-of-mass energy $E$ and the scattering angle $\theta$ are employed.

In total we find that
\begin{multline}
\frac{d \sigma}{d \Omega} =\frac{G_N(E)^2}{E^2}2^{-4-d} \left(E^2\right)^{d/2}  \pi ^{4-d} (7+\cos[2 \theta ])
\csc[\theta ]^2 
 \Bigg(-8 (d-2)^2\\
+\Big(-2688-2720 d+3640 d^2  
-1396 d^3+262 d^4-25 d^5+d^6\Big)
\csc[\theta ]^2 \\
+64 (d-2)^2 (5+3 \cos[2 \theta ]) \csc[\theta ]^6\Bigg) \left(\sin[\theta ]^2\right)^{d/2}  \ ,
\end{multline}
which reduces to \Eqref{Gupta} for $d=4$.

\section{Summing over Kaluza-Klein graviton modes in asymptotically safe quantum gravity \label{sec:app_mass}}

When evaluating the integral over the KK states in the $s$ channel, one has to be careful as one encounters a singularity in the propagator $(s-m^2+i\epsilon)^{-1}$, 
which corresponds to real graviton production at  $s=m^2$.
The $i\epsilon$-prescription can be dealt with following the procedure as outlined in \cite{Han:1998sg}. 
However, here we point out a mistake in equation (B11) in \cite{Han:1998sg} and thus sketch the most important steps in the following for clarity. 
As common, we rewrite the discrete sum over Kaluza-Klein masses into an integral, since for the energies under consideration the spacing 
between adjacent Kaluza-Klein modes is negligible. We use that
\begin{eqnarray}
\frac{1}{M_{\rm Pl}^2}\sum_m&=& \frac{1}{M_{\ast}^{n+2}} \left(2 \pi r \right)^{-n}\sum_m
=\frac{1}{M_{\ast}^{n+2}}\frac{1}{(2 \pi)^n} \int {\rm d}^n m \nonumber\\
&=& \frac{1}{M_{\ast}^{n+2}} \frac{1}{(2 \pi)^n}2 \pi^{\frac{n}{2}}\frac{1}{\Gamma\left( \frac{n}{2}\right)} \int {\rm d}m\,  m^{n-1}.\label{KKmass1}
\end{eqnarray}
Next, we perform the integration over the KK masses $m$. 
We thus have that
\begin{equation}
\int {\rm d}m \, m^{n-1} \frac{i}{s-m^2+ i\epsilon}
= i \left(-i \pi+ \mathcal{P}\int {\rm d}m \, m^{n-1} \frac{1}{s-m^2} \right),
\end{equation}
where $\mathcal{P}$ denotes the principal part of the integral.

In the following, we denote by $A[k_{\rm trans}]$ the principal part of the integral.
We have the following integral in the $s$ channel: 
\begin{eqnarray}
A[k_{\rm trans}]&=& \int_0^{k_{\rm trans}}{\rm d}m \frac{m^{n-1}}{s-m^2}= \frac{s^{\frac{n}{2}-1}}{2} \int_0^{\frac{k_{\rm trans}^2}{s}}{\rm d}x \frac{x^{\frac{n}{2}-1}}{1-x}\nonumber\\
&=&
\begin{cases}
\frac{s^{\frac{n}{2}-1}}{2}\Bigl(-\ln \left( \frac{k_{\rm trans}^2}{s}-1\right) - \sum_{l=1}^{\frac{n}{2}-1} \frac{1}{l}
\left( \frac{k_{\rm trans}^2}{s} \right)^l\Bigr) \,  & \mbox{for n even}\\
\frac{s^{\frac{n}{2}-1}}{2}\Bigl(\ln \left( \frac{1+\sqrt{\frac{k_{\rm trans}^2}{s}}}{\sqrt{\frac{k_{\rm trans}^2}{s}}-1}\right) 
- \sum_{l=1}^{n-2} \frac{1}{l/2} \left(\frac{k_{\rm trans}^2}{s} \right)^{\frac{l}{2}}\Bigr) \, &
 \mbox{for n odd}\end{cases}
\label{eq:lowerint}
\end{eqnarray}

The $t$ and $u$ channel summation over Kaluza-Klein modes does not have a pole, so we can straightforwardly perform the full integral to obtain:
\begin{eqnarray}
&\phantom{=}& \int_0^{k_{\rm trans}} {\rm d}m \frac{m^{n-1}}{-m^2 - \vert t \vert}= -\frac{\vert t \vert^\frac{n-2}{2}}{2}\int_0^{\frac{k_{\rm trans}^2}{\vert t \vert}}{\rm d}x
\frac{x^{\frac{n-2}{2}}}{1+x}\label{eq:upperintint}\\
&=&
\begin{cases}
-\frac{1}{2}\vert t \vert^\frac{n-2}{2} (-)^{\frac{n}{2}-1}\Bigl(\ln \left( 1+\frac{k_{\rm trans}^2}{\vert t \vert}\right)+ \sum_{l=1}^{\frac{n}{2}-1}
\frac{ (-)^l}{l}\left(\frac{k_{\rm trans}^2}{\vert t \vert}\right)^l\Bigr)\,  & \mbox{for n even}\\
 -\frac{1}{2}\vert t \vert^\frac{n-2}{2} (-)^{\frac{n-1}{2}} \Bigl( 2 \tan^{-1}\left(\sqrt{\frac{k_{\rm trans}^2}{\vert t \vert}} \right) 
+\sum_{l=1}^{\frac{n-1}{2}}\frac{2\,(-)^l}{2l-1} \left( \frac{k_{\rm trans}^2}{\vert t \vert}\right)^{\frac{2l-1}{2}}\Bigr)\, &
 \mbox{for n odd}\end{cases}
\nonumber
\end{eqnarray}
The same equations hold if the integral is evaluated at $t \rightarrow u$.
Note that whereas \Eqref{eq:lowerint} is in agreement with equation (B8) in  \cite{Han:1998sg}, our result of \Eqref{eq:upperintint} does not agree with equation (B11) of  \cite{Han:1998sg}.

Let us now analyse the contribution from the upper part of the integral over KK modes, which is dominated by fixed-point scaling behaviour.
Here we have no pole in any of the integrals, since we work under the assumption $s < k_{\rm trans}$, which holds very well in the experimental setting of interest. 
Thus the integrals can be performed straightforwardly and yield:
\begin{eqnarray}
&{}&\int_{k_{\rm trans}}^{\infty} {\rm d}m \frac{m^{n-1}}{(s-m^2)m^{n+2}}=\frac{1}{s^2}\int_{\frac{k_{\rm trans}^2}{s}}^{\infty}{\rm d}x \frac{1}{x^2(1-x)}\nonumber\\
&=&\frac{1}{s^2}\left( \frac{s}{k_{\rm trans}^2}+ \ln \left( \frac{k_{\rm trans}^2/s-1}{k_{\rm trans}^2/s}\right)\right).
\end{eqnarray}
In the $t$-channel we have that
\begin{eqnarray}
&{}&\int_{k_{\rm trans}}^{\infty}{\rm d}m\frac{-m^{n-1}}{\left(\vert t \vert +m^2 \right)}m^{n+2}=- 
\frac{1}{\vert t \vert ^2}\int_{\frac{k_{\rm trans}^2}{\vert t \vert}}^{\infty}{\rm d}x \frac{1}{x^2(1+x)}\nonumber\\
&=& - \frac{1}{\vert t \vert^2} \left( \frac{\vert t \vert}{k_{\rm trans}^2}+ \ln \left(\frac{k_{\rm trans}^2/\vert t \vert}{1+k_{\rm trans}^2/\vert t \vert} \right)\right)
\end{eqnarray}
and similarly in the $u$-channel.

\end{document}